\documentclass[12pt]{article}
\usepackage{amssymb}
\usepackage{amscd}
\renewcommand{\baselinestretch}{2.0}
\newcommand {\newsection}{\setcounter{equation}{0}\section}

\newcommand{\be}{\begin{equation}}
\newcommand{\ee}{\end{equation}}
\newcommand{\ba}{\begin{eqnarray}}
\newcommand{\ea}{\end{eqnarray}}
\newcommand{\bea}{\begin{array}}

\def \rr  {{\mathbb R}}

\def\rep{representation}
\def\trans{transformation}
\def\renorme{renormalizable}
\def\renormy{renormalizability}
\def\hyper{hyper-charge}
\def\lh{left-handed}
\def\rh{right-handed}
\def\nc{noncommutative }
\def\ncy{noncommutativity}
\def\com{commutative}
\def\sm{Standard Model}
\def\Uo{$U_{\star}(1)$}
\def\Ut{$U_{\star}(2)$}
\def\Uth{$U_{\star}(3)$}
\def\Uott{$U_{\star}(3)\times U_{\star}(2)\times U_{\star}(1)$}
\def\Un{$U_{\star}(n)$}
\def\uo{$u_{\star}(1)$}
\def\ut{$u_{\star}(2)$}
\def\uth{$u_{\star}(3)$}
\def\un{$u_{\star}(n)$}
\def\sp{$\star$-product}

\def\nbyn {$n\times n$}

\def\af{anti-fundamental}
\def\fun{fundamental}

\begin{document}

\thispagestyle{empty} \setcounter{page}0
\renewcommand{\baselinestretch}{0.1}
\begin{flushright}
hep-th/0107055\\[-0.4cm]
HIP-2001-25/TH\\[-0.4cm]
IPM/P-2001/024\\[-0.4cm]
\today
\end{flushright}
\renewcommand{\baselinestretch}{1.2}
\begin{center}
{\Large \bf{Noncommutative Standard Model:\\
Model Building}}
\vskip 1cm
{\bf M. Chaichian$^{\dagger}$,
P. Pre\v{s}najder$^{\dagger,a}$, M. M. Sheikh-Jabbari$^{\dagger,b,c}$
\ \ and \ \ {{A. Tureanu}}$^{\dagger}$}
\vskip 0.2cm
{\it $^{\dagger}$High Energy Physics Division, Department of
Physics,
University of Helsinki\\
\ \ {and}\\
\ \ Helsinki Institute of Physics,
P.O. Box 64, FIN-00014 Helsinki, Finland\\
$^a$Department of Theoretical Physics, Comenius University, Mlynsk\'{a} dolina, SK-84248 Bratislava,
Slovakia \\
$^{b}$ The Abdus Salam ICTP,
Strada Costiera 11,Trieste, Italy\\
$^c$Institute for Studies in Theoretical Physics and Mathematics\\[-0.4cm]
P.O. Box 19395-5531, Tehran, Iran}
\end{center}
\vskip 1cm

\begin{abstract}
A \nc version of the usual electro-weak theory is constructed. We
discuss how to overcome the two major problems: 1) although we can have
\nc $U(n)$ (which we denote by $U_\star(n)$) gauge theory we cannot have
\nc $SU(n)$
and 2) the charges in \nc QED are quantized to just $0, \pm
1$.   We show how the latter problem with charge quantization, as
well as with the gauge group, can be resolved by taking $U_{\star}(3)\times
U_{\star}(2)\times U_{\star}(1)$ gauge group and reducing the extra
$U(1)$ factors in an appropriate way.
Then we proceed with building the \nc version of the standard model by specifying the
proper representations for the entire particle content of the theory,
the gauge bosons, the fermions and Higgs. We also present the full action
for the \nc \sm\ (NCSM). In addition, among several
peculiar features of our model, we address the {\it inherent} CP
violation and new neutrino interactions.
\end{abstract}
\newpage

\renewcommand{\baselinestretch}{2.0}
\baselineskip 0.65 cm
\newsection{Introduction}
Undoubtedly, the usual particle physics \sm\ is among the most successful physical theories
and so far it has passed all the precision tests and is capable of
explaining all the present data, or those phenomena and concepts which can be accommodated
within its mathematical structure, such as quarks and neutrino mass and mixing.
The only unobserved, or perhaps theoretically  less elegant, part is the Higgs sector.

Although being experimentally so successful, perhaps its only weak point is the large
number of theoretically undetermined parameters. Mainly motivated by this point, there has
been a lot of work devoted  to formulating theories beyond \sm, through which one can find
some relations between the parameters of the \sm\ and in this way reduce the number of free
parameters. Among these very different attempts  one can mention the grand unified theories
(GUT's) and the minimal supersymmetric \sm\ (MSSM).

In this work we construct a model beyond the \sm\ from a completely different perspective, i.e. the
\sm\ on a \nc space-time, the \nc\sm\ (NCSM).
Noncommutative space-time can be presented by the so-called
Moyal plane, with the coordinates and their conjugate
momentum operators, $\hat{x}_\mu,\ \hat{p}_\nu$, satisfying
\ba
[\hat{x}_\mu,\hat{x}_\nu]=i\theta_{\mu\nu}\ &,& \ \ \ \theta_{\mu\nu}=-\theta_{\nu\mu}\ ,\cr 
[\hat{x}_\mu,\hat{p}_\nu]=i\hbar\eta_{\mu\nu}\ &,& \ \ \
[\hat{p}_\mu,\hat{p}_\nu]=0\ .
\ea
In the above, $\theta_{\mu\nu}$, the \ncy\ parameter (usually taken as a constant tensor), 
is of dimension of $(length)^2$. As it
is seen, the Lorentz symmetry is lost, but, we expect to find the manifest Lorentz
symmetry at low energies, $E^2\theta\ll 1$ (at least if we ignore the quantum
corrections), where $\theta$ is the dimensionful scale of the $\theta_{\mu\nu}$ tensor. 
Then, one should define field theory on the \nc space-times, \nc field
theory. To pass to \nc field theories, it is enough to replace the usual product of the
fields in the (\com) action, by the Moyal \sp\footnote{We note that this recipe cannot be
used for gauge theories other than \Un.}
\begin{eqnarray}\label{star}
(f\star g)(x)&=&{\rm e}^{{i\over 2}\theta_{\mu\nu}
\partial_{x_{\mu}}\partial_{y_{\nu}}}f(x)g(y)\Big|_{x=y}\cr
&=&f(x)g(x)+{i\over2}\theta_{\mu\nu}
\partial_{{\mu}}f\partial_{{\nu}}g+{\cal{O}}({\theta^2})\ .
\end{eqnarray} 
Introducing this \sp\ into the actions has some non-trivial consequences both at
the classical (tree) and quantum (loop) levels. 

At classical level, among these consequences, we would like to mention the restrictions it
imposes on the gauge theories: only the \nc $U(n)$ gauge theories have a simple \nc
extension and we cannot even have \nc $SU(n)$ gauge theories. Furthermore,
the \rep s for the \un\ algebra are restricted to those of \nbyn\ hermitian matrices
\cite{nogo}. Also, \ncy\
imposes severe restrictions on the fermions and their charges \cite{{nogo},{Haya}}.
We shall discuss these points in more detail in the next section. The other interesting
classical consequence of \ncy\ is the inherent $C$ and $CP$ violation in the \nc field
theories \cite{CPT}.

As for the quantum level, we can mention the loop calculations and \renormy\ discussions. 
During the past two years there has been a large number of articles on that subject  
 (see, e.g.,
\cite{Filk}-\cite{Loriano})
\footnote{For a string theory survey on \nc issues, see \cite{SW}.}. From all
these results here we mention only two:
\newline
{\it i)} In general, the unitarity of \nc field theories is
related to having a space-like \ncy, i.e. $\theta_{\mu\nu}\theta^{\mu\rho}$ as a matrix should
be positive definite \cite{Unitary};
\newline
{\it ii)}  An intrinsic and general feature of the \nc field theories is the so-called
IR/UV mixing \cite{Sei1}: although we can usually remove the UV divergences in the \nc
version of the usual \com\ \renorme\ theories by adding proper counter-terms (and hence the
theory is UV
\renorme), upon sending the UV cut-off to infinity we remain with some new IR divergences.
There have been three proposals to resolve this IR divergence problem
\cite{{Sei1},{Sei2}};\cite{Hard};\cite{Wien}, among which are  
the \nc hard resummassion \cite{Hard}, and/or introducing a new way of regularization
\cite{Wien}; we believe that, one way or another, this problem can be
removed.

In particular we would like to point out that the \nc gauge theories
\cite{{Carmello},{Loriano}}, the
\nc version of real $\phi^4$ theory \cite{{Sei1},{Amicu},{Roiban}} as well as the complex
$\phi^4$ theory \cite{Aref} and the \nc version of QED (NCQED) \cite{{Haya},{Ihab}} have
been shown to be one-loop \renorme.

There have also been many attempts to study the phenomenological
consequences of \nc field theories (by taking the space-time to be a \nc
Moyal plane) \footnote{Noncommutative geometry (in a general sense) has been
previously used to build a theory beyond \\[-.3cm] Standard Model, see e.g.,
\cite{NCG}. 
Recently within the Connes formulation, the unimodularity condition have
\\[-.3cm] been used to
obtain the hyper-chrages for the fermions \cite{Lazzarini}.  However, these models
are based on a very \\[-.3cm] different approach than ours, where the fields evolve
in almost
commutative spaces (the space-time is \\[-.3cm] commutative with a minimal
noncommutativity in the internal space).}. 
However, most of them are aimed at accommodating the extra
\nc contributions within the error bars of the present data
\cite{Zdecay}-\cite{Banks}. A rigorous
and robust mathematical framework which is not suffering from the charge quantization
problem \cite{{nogo},{Haya}} and the extra $U(1)$ factors (in the \Un\ gauge 
theory
compared to $SU(n)$) \cite{{Armoni},{Loriano}} is not yet constructed. This is exactly what
we would like to do in this paper. We will show how, by fixing the gauge group of the \nc\sm\
(NCSM) to \Uth$\times$\Ut$\times$\Uo\ and reducing the two extra $U(1)$ factors 
through the
appropriate Higgs particles, the number
of possible particles in each family (which are six: left-handed leptons, right-handed
charged lepton, left-handed quarks, right-handed up quarks, right-handed down quarks and
Higgs) is fixed, as well as their hyper-charges (and hence the electric charge). We would like to
emphasize that the existence of the Higgs particle, in our model, is an unavoidable outcome. 
As a consequence, 
two extra massive gauge bosons and two extra massive scalar particles will appear.

{\it Convention:}

In order to make a distinction between these two types of scalar fields which we have:
the one(s) which we use for the {\it reduction} of the extra 
$U(1)$ symmetries and the usual \sm\
Higgs, which is used for breaking the electro-weak symmetry, we call the former one as
``{\it Higgsac}'' and
keep the ``Higgs'' for the usual Higgs doublet
\footnote{The suffix ``ac'' stems either 
from the word ``acommutative'' (i.e. not commutative) or from the diminutive 
suffix in Persian,
similar to ``ino'' in Italian, and hence ``Higgsac'' is equivalent to small 
Higgs. We use this
terminology to distinguish these scalars from the usual 
Higgs and also the Higgsinos of MSSM.}. 

The paper is organized as follows. In section 2, we review the problems
and restrictions for constructing a \nc version of \sm\ and discuss a mechanism
to resolve these problems. 
In section 3, in order to show how our procedure works in practice, we work out the details
of the reduction of  the extra $U(1)$ factor(s) and show how this also resolves 
the
charge quantization problem, for the particular case of the \nc version of QCD+QED
which can be denoted  by $NC(SU_c(3)\times U(1))$ gauge theory. 
In section 4, which in a sense is the main part of the paper,  
we construct  the NCSM. We start with the \Uth$\times$\Ut$\times$\Uo\
gauge theory and reduce the two extra $U(1)$ factors by introducing two extra 
Higgsac
particles in proper \rep s. Then, we proceed with introducing matter fields and discuss
in detail how the hyper-charges are fixed to those of the usual \sm. 
In section 5, we work out the
details of the electro-weak symmetry breaking. In this way we define the photon, $Z$ and $W^\pm$
fields. Then, in the fermionic part, we discuss the interaction terms for the fermions and compare
them with the usual \sm\ as well as the corresponding Yukawa couplings and mass terms. 
In section 6, among several new features of NCSM,
we mention the neutrino dipole moment and the \nc correction to the weak-mixing angle,
$\theta_W$, or more precisely to the $\rho$ parameter and ${m^2_W\over m^2_Z}$ ratio. 
In this way we impose some upper bounds on the masses of two extra massive gauge
boson as well as on the \ncy\ parameter. Finally in section 7, we discuss some of the open
questions. More detailed analysis of the normal sub-groups of \Un\ as well as 
the Higgsac symmetry reduction is gathered in the Appendices.

\newsection{The major problems in constructing NCSM and \\ [- 0.6cm] the
proposal to resolve them}

In this section we recapitulate the problems one encounters in building a \nc version of the \sm\
and present the way out of them.
These problems and restrictions which we classify in three sets, are all imposed by
the mathematical (group theoretical) structure of \nc gauge theories.
However, first let us review some related information about the usual \sm. The usual \sm\ in
the gauge bosons sector contains 8 (massless) gluons, 1 (massless) photon and 3
(massive) weak gauge bosons. We have integrated the information about the matter fields and
their charges in the following table.

\begin{table}[ht]\label{tab:sc}
\vspace{0.3cm}
\begin{center}
\begin{tabular}{|c|c|c|c|c|}
\hline
Particles& Electric charge & $SU(2)$ weak charge& Hyper-charge& Colour charge \\
\hline
LH electron& $-1$ & $-{1\over 2}$ & $-1$ & none \\
LH neutrino& $0$  & $+{1\over 2}$ & $-1$ & none\\
RH electron& $-1$ & $0$         & $-2$   & none\\
\hline
LH up quark& $+{2\over 3}$ & $+{1\over 2}$ & $+{1\over 3}$ & has \\
LH down quark&$-{1\over 3}$& $-{1\over 2}$ & $+{1\over 3}$ & has \\
RH up quark& $+{2\over 3}$ & $0$           & $+{4\over 3}$ & has \\
RH down quark&$-{1\over 3}$& $0$           & $-{2\over 3}$ & has \\
\hline
Higgs      & $0$ & $-{1\over 2}$ & $+1$ & none\\
\hline
\end{tabular}
\end{center}
\caption{LH=left handed, RH=right handed}
\end{table}

Now we are ready to discuss the three major problems.

\subsection{Problems}

{\it i) Charge quantization problem}:

As it was shown in \cite{Haya}, the charges for the matter fields coupled to the \Uo\ theory must be
quantized to just $0,\pm 1$, depending on the \rep\ of particles. This is due to the fact
that in a sense the \Uo\ theory is a non-Abelian theory (for a more detailed discussion we refer
to \cite{{Haya},{nogo}}). Now, we face the first and the most challenging obstacle:
As we explicitly see from the table,
not all the electric or hyper-charges of the particles fulfill this
condition. So, not only we are not able to construct NCQED, but going
to the electro-weak level (and considering the \hyper s) makes the problem worse and we
face a larger variety of non-quantized \hyper s.
\newline
{\it ii) The extra gauge fields}:

According to \nc group theoretical arguments (e.g. see \cite{nogo}), the \Uo\ sub-group of \Un\
is not a normal sub-group and therefore mathematically it is not possible to define a \nc
$SU(n)$ algebra (or group) by simple insertion of $\star$-products.
However, even if we ignore this mathematical fact and drop
the corresponding \Uo\ gauge field in the \Un\ gauge theory action, the remaining
theory is not \renorme\ \cite{{Armoni},{Loriano}}. Consequently, as a direct generalization
of the $SU_c(3)\times SU_L(2)\times U(1)$ gauge theory, one cannot avoid two extra
$U(1)$ factors, i.e., two extra gauge fields appearing in NCSM.
\newline
{\it iii) The no-go theorem}:

In \cite{nogo}, based on group theoretical arguments, we have proved a no-go theorem stating
that:
\newline
a) the local \un\ {\it algebra} only admits the irreducible \nbyn\
matrix-representation. Hence the gauge fields are in \nbyn\ matrix
form, while the matter fields {\it can only be} in fundamental, adjoint or singlet states
\footnote{Within superfield approach similar arguments have been presented in
\cite{Teras}.};
\newline
b) for any gauge group consisting of several simple-group factors, the matter fields can
transform nontrivially under {\it at most two} \nc group factors. In other words, the matter
fields cannot carry more than two \nc gauge group charges.

The a) restriction is actually what we have already had in the usual \sm, i.e. all the gauge
bosons as well as the matter fields are sitting in the \rep s which are also allowed in the
\nc case. However, as for the b) criterion, it is clear from the table that the
particles coupled to gluons, the quarks, carry {\it three} different charges,
i.e. \hyper, weak $SU(2)$ charge and colour charge.

Before explaining our procedure to resolve the above mentioned problems,
however, we would like to make a comment on the no-go theorem.
The arguments of \cite{nogo}, and in particular part b), are based on the invariance of the action
under the {\it finite} gauge transformations.
In other words, to define the gauge transformation for the matter fields we have considered
the {\it group} factors, while in principle it is also possible to define these gauge
transformations only with the {\it algebra} (i.e. the infinitesimal gauge transformations),
in which case one can relax the condition b) \footnote{We would like to thank L. Bonora for a
discussion on this point.}. For the usual Lie-groups and algebras
where the group elements are obtained through the simple exponentiation of the algebra
elements, of course the infinitesimal and finite gauge transformations are resulting in the
same physics (at least for Yang-Mills theories).
However, this is not always the case, a famous example being the Chern-Simons theories in
which, although the theory is invariant under infinitesimal gauge transformations, the
invariance under finite gauge transformations is not immediate. As a result, to have a
well-defined quantum Chern-Simons theory, the level should be quantized, which  in
turn is  an implication of finite gauge transformations.
For the \nc groups when the gauge group involves more than one simple \Un\ factor, the
relation between the algebra and the corresponding group is not given by a simple
star-exponentiation \cite{nogo}. We believe that it is the invariance under
the {\it finite} gauge transformations which is indeed fundamental, and of course this also
covers the infinitesimal gauge invariance.

\subsection{The way out}

To show the way out of the above mentioned problems we recall two facts:
\newline
{\it I)} In the usual physical models,  there is always a $U(1)$ factor together with the
$SU(n)$ factors, i.e. $SU_c(3)\times U_q(1)$ for QCD+QED and $SU_c(3)\times
SU_L(2)\times U_Y(1)$  for the \sm.
\newline
{\it II)} If we define the photon (or the hyper-photon) through a linear combination of two
(or three) \Uo\ fields, although the charge for each \Uo\ factor is quantized restrictively to 0 and $\pm1$,
there is the
chance to find more variety of charges (but still quantized). Furthermore, this shows a way out
of the implications of part b) of our no-go theorem.

Hopefully there is a standard and well-known procedure to
implement the above two facts: the Higgs symmetry breaking
scenario. Hence our recipe is to start with \Uth$\times$\Uo\ (or
\Uott) and {\it reduce} two (or three) $U(1)$ factors to {\it one}
$U(1)$ factor,  through one (or two) proper Higgsac field(s). We
would like to emphasize that in order to reduce a symmetry through
the Higgs mechanism it is necessary that the Higgs is in a
non-singlet \rep\ of that symmetry. Therefore, in our case, the
Higgsac field(s) should be charged {\it only} under the $U(1)$
sub-group of \Uth\ (and \Ut) as well as under the individual \Uo.
Indeed the \Un\ group enjoys the
property of having the needed $U(1)$ normal sub-group, which here it 
is denoted by $U_n(1)$. (For the
definition of $U_n(1)$ sub-group see Appendix A.)

In sections 3 and 4, we will explicitly and in details show how the above
observation works and how tightly it fits into the existing matter content of the \sm.

\newsection{The \nc QED+QCD}

To build the \nc version of the $SU_c(3)\times U(1)$ gauge theory,
first we need to introduce the gauge group which will be denoted
by $NC(SU(3)\times U(1))$. In order to achieve this goal and
clarify the notation we need to describe the structure of the
group $U_\star (n)$ in more detail. The $U_\star (n)$ stands for
the usual noncommutative version of $U(n)$ obtained by insertion
of the $\star$-product between the $U(n)$ matrix valued functions.
Consequently, all $U_\star (n)$ matrix elements are power series
in $\theta$. Taking this into account, the $U_\star (n)$ has two
invariant (normal) sub-groups: 1) The group $NCSU(n)$ obtained
from $\star$-product of $SU(n)$ matrix valued functions (which do
contain $U(1)$ part, however, at least linear in $\theta$ so that
in the limit $\theta\to 0$ it reduces to the usual gauge group
$SU(n)$). Therefore one can define the {\it factor}-group
$U_n(1)=U_{\star}(n)/NCSU(n)$. Note that $U_n(1)$ is a commutative
Abelian sub-group; and 2) the $U^n_\star (1)$ sub-group, obtained
by the action of $U_\star (n)$ on its $U_\star(1)$ sub-group. This
$U_{\star}(1)$ sub-group is generated by star exponentiation of
the trace of $u_\star(n)$ algebra elements, i.e.
$exp_{\star}(iTr\lambda){\bf 1}_{n\times n}, \ \ \lambda\in
u_\star (n)$, where the trace is taken over the $n\times n$
matrices. More explicitly, if $h\in U_{\star}(1)$ sub-group and
$g\in U_{\star}(n)$, then the elements of $U^n_{\star}(1)$ are of
the form of $g\star h\star g^{-1}$. We stress that this $U_\star (1)$
is not an invariant sub-group whereas $U^n_\star (1)$ is; and
we  emphasize that it is the factor-group $U_n(1)$ which is 
used in our standard model construction, while the other invariant sub-group,  
$U^n_\star (1)$, is not used through out the paper.
Also note that both of the $NCSU(n)$ and $U^n_{\star}(1)$ sub-groups
should be understood as power series in $\theta$. 
The details of the sub-group construction are given in Appendix A.

To obtain the NC(QED+QCD) we start with
the \Uth$\times$\Uo\ gauge theory, establish the particle content
and the representations, give the gauge transformations and write the
gauge-invariant action.
Subsequently, by a properly chosen Higgsac boson, we reduce
the two existing $U(1)$ factors to a single $U(1)$ gauge symmetry, or more
precisely the gauge group is reduced to $NC(SU(3)\times U(1))$.
The final $U(1)$ factor will be proven to correspond to \nc
version of QED. Finally, we shall address the new
features and interactions of NC(QED+QCD), like CP violation, new
"multi-photon" interactions and photon-gluon
interactions.

\subsection{The field content of the model; fixing the conventions}

In the following, we shall fix our notations and also point out the fact
that the \sp\ will be omitted everywhere from now on, and
unless mentioned explicitly, it
is understood that the \sp\ is there.

The pure \Uth$\times$\Uo\ theory is described by one gauge field, $B_{\mu}$, 
valued in the \uo\ algebra and
the \uth-valued gauge fields:
\be\label{un-g.f.}
G_{\mu}(x)=\sum_{A=0}^{8}\ G_{\mu}^A(x) T^A\ .
\ee
According to \cite{nogo}, the gauge fields corresponding to \uth\ are necessarily in a $3\times3$
matrix form, because no other representation for the \uth\ algebra is possible. As a result, we
can take the generators
$T^a,\ a=1,2,\cdots, 8$ to be the Gell-Mann matrices, while $T^0={\bf 1}_{3\times 3}$.

If we denote the elements of \Uo\ by $v(x)$ and the elements of \Uth\  by $U(x)$, we can write
the finite local transformations of the gauge fields as
\ba\label{g.f.t.}
B_{\mu}&\to& v B_{\mu} v^{-1}  + {i\over g_1} v \partial_{\mu}v^{-1} \cr
G_{\mu}&\to& U G_{\mu} U^{-1}  + {i\over g_3} U\partial_{\mu}U^{-1}\ .
\ea
Then the gauge field strengths
\ba\label{f.s.}
B_{\mu\nu}=\partial_{[\mu}B_{\nu]} + i g_1[B_{\mu},B_{\nu}]_{\star}\ ,\cr
G_{\mu\nu}=\partial_{[\mu}G_{\nu]} + i g_3[G_{\mu},G_{\nu}]_{\star}\ ,
\ea
will transform as $B_{\mu\nu}\to v B_{\mu\nu}v^{-1}\ $ and $\
G_{\mu\nu}\to U G_{\mu\nu}U^{-1}$, leaving the
action of the \Uth$\times$\Uo\ Yang-Mills theory
\be\label{G-action}
S_{NCYM}=-{1\over 4}\int d^4x\big[B_{\mu\nu}B^{\mu\nu}+{\rm 
Tr}(G_{\mu\nu}G^{\mu\nu})\big]\ ,
\ee
invariant. A full account of these issues, towards the scope of this paper, is given in
\cite{{nogo},{Loriano}}.

As for the matter content of the \Uth$\times$\Uo\ theory, the number of
independent charged particles that
can occur in this model, according to the no-go
theorem \cite{nogo}, is ${1\over 2}2\times(2+1)=3$, since the number of simple group factors is two. We take these particles
to be the electron (in anti-fundamental representation of \Uo), the up quark (in fundamental
\rep\ of \Uo\ and anti-fundamental \rep\ of \Uth) and the down quark (in
anti-fundamental \rep\ of
\Uth).
Then, the gauge transformation properties of the
fermions under the \Uth$\times$\Uo\ gauge group are:

\ba\label{f.trans}
\psi_e(x) &\rightarrow& \psi_e(x)v^{-1}(x)\ ,\cr
\psi_u(x) &\rightarrow& v(x)\psi_u(x)U^{-1}(x)\ , \cr
\psi_d(x) &\rightarrow& \psi_d(x)U^{-1}(x)\ .
\ea

The gauge invariant action corresponding to this \Uth$\times$\Uo\ model is
\ba\label{action13}
S &=&\int d^4x\big[\bar{\psi_e}\gamma^{\mu}D^1_{\mu}\psi_e+\bar{\psi_u}\gamma^{\mu}D^{1+3}_{\mu}\psi_u
+\bar{\psi_d}\gamma^{\mu}D^3_{\mu}\psi_d \cr
&-&{1\over 4}B^{\mu\nu}B_{\mu\nu}-{1\over 4}Tr (G^{\mu\nu}G_{\mu\nu})\big]\ .
\ea

The covariant derivatives entering (\ref{action13}) are:
\ba
D^{1}_{\mu}&=&\partial_{\mu}-{i\over 2}g_1B_{\mu}\\
D^{3}_{\mu}&=&\partial_{\mu} -{i\over 2}g_3G^{'0}_{\mu}-{i\over 2}g_3G^{a}_{\mu}T^a\\
D^{1+3}_{\mu}&=&\partial_{\mu} +{i\over 2}g_1B_{\mu}-{i\over 2}g_3G^{'0}_{\mu}-{i\over 2}
g_3G^{a}_{\mu}T^a\ .
\ea
For a reason that will become clear momentarily, we have denoted the zeroth component of the
\Uth\ gauge field by $G^{'0}_\mu$.

Still, this is not NC(QCD+QED), but a theory that suffers of the charge quantization problem. In
order to cure it, we use the Higgs procedure for reducing the extra $U(1)$ 
factors of
\Uth$\times$\Uo\ to a
single $U(1)$, which will exhibit the properties of a true \nc version of QED 
in the coupling of
the \nc photon to the
fermionic fields.

The reduction of symmetry has to be done through a proper Higgsac
field, i.e. a scalar particle that is charged under those groups
(or sub-groups) that we intend to reduce. In this case, the scalar
field has to be  charged under the $U_1(1)$ and $U_3(1)$ 
invariant sub-groups of
$U_\star (1)$ and   $U_\star (3)$ factors.
The gauge transformation undergone by
the symmetry-breaking scalar, Higgsac field, is
\be\label{Higgsacg.t.}
\Phi(x)\to U_1(x)\Phi(x)v^{-1}(x)\ , 
\ee 
where $U_1(x)$ is the {\it $\theta$-independent} phase factor of $U_3(1)$ and 
$v(x)\in U_1(1)$ (for more details see Appendix A). 
We should stress that $\Phi(x)$ is $\theta$-independent and in 
Eq.(\ref{Higgsacg.t.}) the usual product (and not 
$\star$-product) should be used.
Since the $NCSU(3)$ and hence $U_3(1)$ sub-groups should be 
understood as power series expansions in $\theta$, the symmetry reduction
problem should be investigated systematically in the same power
series. We stress that the $U_1(1)$ and $U_3(1)$ phase factors are 
$U_\star(3)\times U_\star(1)$ invariant. 
The details of the symmetry reduction are given in the Appendix B.

The only gauge invariant terms introduced in the gauge invariant
action by the presence of the scalar field are:

\be\label{Higgsac-act}
(D^{1+1}_{\mu}\Phi)^{\dagger}(D^{1+1}_{\mu}\Phi)+m^2\Phi^{\dagger}\Phi-{f\over
4!}(\Phi^{\dagger}\Phi)^2
\ee
with the covariant derivative given by:
\be\label{Higgsaccov}
D^{1+1}_{\mu}=\partial_{\mu} +{i\over 2}3g_3G^{'0}_{\mu}-{i\over 2}g_1B_{\mu}\ ,
\ee
where by $G^{'0},\ B$ in the above we only mean the $\theta$-independent
parts of the corresponding gauge fields. These 
$\theta$-independent parts are those which transform properly under 
$U_\star(3)\times U_\star(1)$.  
Note that in the Eqs.(\ref{Higgsac-act}-\ref{Higgsaccov}) the usual product of 
functions should be used.

Applying the usual Higgs mechanism, we shall obtain a massive gauge boson,
$G^0_{\mu}$, whose mass term in
the Lagrangian is:
\be\label{g_mu}
{1\over 4}(3g_3G^{'0}_{\mu}-g_1B_{\mu})^2\phi_0^2 = N^2(G^{0}_{\mu})^2\phi_0^2 ,
\ee
where $N={1\over 2}\sqrt{g_1^2+(3g_3)^2}$ is a normalization factor and $\phi_0$ is the vacuum
expectation
value for the scalar field.
Actually, in order to write (\ref{g_mu}), we have performed a rotation in the $(B_{\mu}, \,
G^{'0}_{\mu})$ plane,
by the angle $\delta_{13}$
\be\label{theta13}
\tan\delta_{13}={g_1\over 3g_3}\ ,
\ee
so that
\ba\label{rotation13'}
G^{0}_{\mu}&=&\cos\delta_{13}G^{'0}_{\mu}-\sin\delta_{13}B_{\mu},\cr
A_{\mu}&=& \sin\delta_{13}G^{'0}_{\mu}+\cos\delta_{13}B_{\mu}\ ,
\ea
where $A_{\mu}$ is the (massless)  \nc photon, i.e. the gauge field of the 
residual $U(1)$ symmetry.
The reciprocal of this rotation is given by:
\ba\label{rotation13}
G^{'0}_{\mu}&=&\cos\delta_{13}G^0_{\mu}+\sin\delta_{13}A_{\mu},\cr
B_{\mu}&=& -\sin\delta_{13}G^0_{\mu}+\cos\delta_{13}A_{\mu}\ .
\ea

As desired the the Lagrangian (3.11) is $U_\star(3)\times U_\star(1)$ gauge 
invariant. However, the Higgsac
field may interact with other matter fields {\it only} indirectly, 
via the
$\theta$-independent parts of the corresponding gauge fields. Here we 
only investigate these effects in the leading order.
Therefore in this leading order the theory should be 
treated as an ``effective theory'' for energies lower than the noncommutativity
scale, which as we will discuss, can be as low as TeV. From this
point of view our model is an effective theory up to the TeV
scale. The calculation of higher $\theta$-corrections
would require a more detailed analysis which is postponed to future works.


\subsection{Reduction of the \Uo\ symmetries: A solution to the charge\\[-0.6cm]
quantization problem}

Now we show that this Higgs mechanism has indeed brought us to the NC(QCD+QED), by curing the
charge quantization problem that plagues the usual \Uo\ gauge theory. To this end, we show
that the fermions of
the \Uth$\times$\Uo\ theory couple to the massless gauge boson of the residual \Uo,
$A_{\mu}$, through the usual electric charges (see table 1).

For the electron, the coupling to $A_{\mu}$ emerges from the first term of (\ref{action13}),
taking into account (\ref{rotation13}):
\ba
\bar{\psi_e}\gamma^{\mu}D^1_{\mu}\psi_e&=&\bar{\psi_e}\gamma^{\mu}\partial_{\mu}\psi_e
- {i\over 2}g_1\bar{\psi_e}\gamma^{\mu}\psi_eB_{\mu} \cr
&=&\bar{\psi_e}\gamma^{\mu}\partial_{\mu}\psi_e
- {i\over 2}g_1\cos\delta_{13}\bar{\psi_e}\gamma^{\mu}\psi_eA_{\mu}+\cdots\ ,
\ea
where the dots indicate the coupling to the massive gauge boson, $G^0_{\mu}$.
We would like to remind the reader that, although it is not shown
explicitly,  the products between the fields are all performed by the
Moyal star product.

As we want the term relevant for the coupling of the electron to the gauge field $A_{\mu}$ to be
proportional to the electric charge of the electron, i.e. $-e$, we define $e$ as
\be\label{conscheck_e}
{1\over 2}g_1\cos\delta_{13}=e\ .
\ee
A similar reasoning for the down quark will give:
\ba
\bar{\psi_d}\gamma^{\mu}D^3_{\mu}\psi_d &=& \bar{\psi_d}\gamma^{\mu}\partial_{\mu}\psi_d
-{i\over 2}g_3\bar{\psi_d}\gamma^{\mu}\psi_dG^{'0}_{\mu}-
{i\over 2}g_3\bar{\psi_d}\gamma^{\mu}\psi_dG^{a}_{\mu}T^a \cr
&=&\bar{\psi_d}\gamma^{\mu}\partial_{\mu}\psi_d
-{i\over 2}g_3\sin\delta_{13}\bar{\psi_d}\gamma^{\mu}\psi_dA_{\mu}-
{i\over 2}g_3\bar{\psi_d}\gamma^{\mu}\psi_dG^{a}_{\mu}T^a+\cdots\ ,
\ea
from which we find the condition
\be\label{conscheck_d}
-{1\over 2}g_3\sin\delta_{13}=q_d\ ,
\ee
where $q_d$ is the electric charge of the down quark. However, using (\ref{theta13}) and
(\ref{conscheck_e}) we find that
\be\label{d-q-charge}
q_d=-{1\over 3}e\ ,
\ee
which is the correct relation.

For the up quark,
\ba
\bar{\psi_u}\gamma^{\mu}D^{1+3}_{\mu}\psi_u &=&
\bar{\psi_u}\gamma^{\mu}\partial_{\mu}\psi_u +{i\over
2}g_1\bar{\psi_u}\gamma^{\mu}B_{\mu}\psi_u \cr
&-&{i\over 2}g_3\bar{\psi_u}\gamma^{\mu}\psi_uG^{'0}_{\mu}-
{i\over 2}g_3\bar{\psi_u}\gamma^{\mu}\psi_uG^{a}_{\mu}T^a\
\ea
and the relevant terms for the coupling with $A_{\mu}$, having in view (\ref{rotation13}), will be:
\ba\label{upquarkint}
{\cal L}_{u-A_{\mu}}&=&{i\over
2}g_1\cos\delta_{13}\bar{\psi_u}\gamma^{\mu}A_{\mu}\psi_u
-{i\over 2}g_3\sin\delta_{13}\bar{\psi_u}\gamma^{\mu}\psi_uA_{\mu}\cr
 &=&{i\over 2}(g_1\cos\delta_{13}-g_3\sin\delta_{13})\bar{\psi_u}\gamma^{\mu}A_{\mu}\psi_u
-{i\over 2}g_3\sin\delta_{13}\bar{\psi_u}\gamma^{\mu}[\psi_u,A_{\mu}]_{\star}\ ,
\ea
and therefore
\be\label{conscheck_u}
{1\over 2}(g_1\cos\delta_{13}-g_3\sin\delta_{13})=q_u\ .
\ee
Upon using the definition of $e$ (\ref{conscheck_e}) and (\ref{theta13}), we find
\be
q_u=+{2\over 3} e\ . \ee As we see, the charges for the up and
down quarks have come out of the mathematical structure of our
model and they have {\it not} been put by hand. In fact, the only
allowed (possible) charges for the particles which also couple to
the gluons are ${1\over 3}$ and ${2\over 3}$ in units of electron
charge. In other words, the \rep\ fixes completely the electric
charges. The reader may find some more details on the symmetry
reduction in the fermionic sector in Appendix C.

\subsection{Discussions on the model; some new features}

Although we do not tend to analyze the $NC(SU_c(3)\times U(1))$ model described
previously in detail, we would like to point out some of the important consequences and
a more detailed survey is postponed to future  works.
\vskip .5cm
{\it 1) The  \renormy}
\vskip .5cm
Noting the fact that in order to construct our model we started with a \Uth$\times$\Uo\
gauge
theory plus all possible charged matter fields, this theory is (UV) \renorme\
\cite{{Haya},{Loriano}}. In addition we have used a (complex) 
scalar field coupled to the two commutative
$U(1)$ factors with a $(\phi^{\dagger}\phi)^2$ potential and it is well-known 
that this scalar theory is renormalizable.
On the other hand, it is well known that the Higgs scenario does not spoil 
the \renormy\ of the theory. Hence, altogether we expect our theory to be 
\renorme.
\vskip .5cm
{\it 2) The  photon-photon and photon-gluon interactions}
\vskip .5cm
Having the definition of physical fields, $A_\mu,\ G^0_\mu$, in terms of $B_\mu,\ G^{'0}_\mu$,
we can easily read off the interaction of photon with itself and also with other gauge
bosons. Inserting (\ref{rotation13}) into  the action (\ref{G-action}), there are some immediate
results:
\newline
{\it i)} The three- and four-photon vertices are {\it not} exactly what is dictated by a simple
\Uo\ theory. The coefficient (coupling) for the $A_\mu$-$A_\mu$-$A_\mu$ term is
${2\over 3}e(1+2\sin^2\delta_{13})$, while for the $A_\mu$-$A_\mu$-$A_\mu$-$A_\mu$ term it is
${4\over 3}e^2(1+2\sin^2\delta_{13})$;
\newline
{\it ii)} There are $A_\mu$-$G^0_\mu$-$G^0_\mu$,  $A_\mu$-$A_\mu$-$G^0_\mu$ and
$A_\mu$-$A_\mu$-$G^0_\mu$-$G^0_\mu$ interaction terms;
\newline
{\it iii)} The usual gluons (the $G^a_\mu,\ a=1,\cdots,8$ fields) also couple to the photon, $A_\mu$.

As a side effect of the above arguments it is likely that they show a way out of the standing
problem of a simple \Uo\ gauge theory: the negative $\beta$-function \cite{Haya}.
It is an experimentally confirmed fact that the QED coupling, $\alpha$,
increases as we increase the
energy:
\be
\alpha|_{_{E\sim 10eV}}\simeq{1\over 137.036}\ ,\ \ \ \
\alpha|_{_{E\sim m_Z}}\simeq{1\over 128.9}\ .
\ee
On the other hand, a direct one loop calculation for the simple \Uo\ gauge theory shows a negative
$\beta$-function. However, according to our arguments one should keep in mind that
in the $NC(SU_c(3)\times U(1))$ model discussed above, photon is also involved in some
interactions
other than those of the \Uo\ theory. Also, the number of charged particles coupled to the photon is now
increased, as the charge quantization problem of the quarks has been
eliminated. This may show a way to resolve the negative $\beta$-function
problem.
\vskip .5cm
{\it 3) The  fermionic interactions}
\vskip .5cm
Here we would like only to mention about the inherent CP violation because of the \sp\ present in
the fermion-photon coupling terms. As discussed in \cite{CPT}, it is important that the photon
appears on the right-hand-side (or left-hand-side) of the $\psi$ field, like up quark (or electron
and down quark). Consequently, the anti-particle of the up quark (which carries $-{2\over 3}e$
charge) would be coupled to photon form the left-hand-side. More intuitively, the \nc particles,
besides the usual electric charges, also carry higher-pole (including dipole) moments
\cite{{Ren},{Ihab}}; the anti-particle of any particle, not only should carry the opposite
charge, but also the opposite  dipole moment. Since these dipole moments are
proportional to momentum \cite{{Ren},{Ihab}}, the theory would not be CP invariant, while CPT is
conserved \cite{CPT}.

Finally, we would like to note that in the up quark-photon interaction term (\ref{upquarkint}),
besides the usual ${\bar\psi}\gamma^\mu\psi A_\mu$ term, there is a Moyal bracket term which is
not there for electron and down quark. Group theoretically, this is related to the fact that the up quark
carries two different charges while the electron and down quark carry only one type of charge.

\newsection{The \nc \sm\ (NCSM)}
Having worked out the details of the \Uth$\times$\Uo\ gauge theory, the symmetry
reduction scenario
and the charges of the particles as a warm up, we are now ready to present our formulation of
NCSM. In this section, applying the same machinery, but for the group \Uott\ we construct the
NCSM. First we show the reduction of three $U(1)$ factors to the {\it 
hyper-charge} $U(1)$ and
discuss that, as a result, two of the corresponding $U(1)$ fields become 
massive.
Then, we proceed with the matter fields and show that their hyper-charges are fixed to those of
the usual \sm\  (given in table 1).

\subsection{The gauge group}

The pure \Uth$\times$\Ut$\times$\Uo\ theory is described by one gauge field, $B_{\mu}$,
valued in the \uo\ algebra, the \ut-valued gauge fields:
\be\label{u2-g.f.}
W_{\mu}(x)=\sum_{I=0}^{3}\ W_{\mu}^I(x) \sigma^I\
\ee
and
the \uth-valued gauge fields:
\be\label{u3-g.f.}
G_{\mu}(x)=\sum_{A=0}^{8}\ G_{\mu}^A(x) T^A\ .
\ee
For a similar reason as in the previous section, i.e. according to the no-go theorem \cite{nogo},
we take the generators of the \ut\ algebra as the
Pauli matrices  $\sigma^i,\ i=1,2,3$ and $\sigma^0={\bf 1}_{2\times 2}$, while the generators of
the \uth\ algebra will be taken as the Gell-Mann matrices
$T^a,\ a=1,2,\cdots, 8$ and $T^0={\bf 1}_{3\times 3}$.

In the following we continue to denote the elements of \Uo\ by $v(x)$ and the elements of
\Uth\  by $U(x)$, while the elements of \Ut\ are denoted by  $V(x)$.
The local transformations of the gauge fields are of a similar form with (\ref{g.f.t.}) and the action
\be\label{G123-action}
S_{gauge\ fields}=-{1\over 4}\int d^4x\big[B_{\mu\nu}B^{\mu\nu}+
{\rm Tr}(W_{\mu\nu}W^{\mu\nu})+{\rm Tr}(G_{\mu\nu}G^{\mu\nu})\big]\ ,
\ee
is gauge invariant.

In order to reduce the three $U(1)$ factors of the \Uth$\times$\Ut$\times$\Uo\ 
theory we
should use
two scalar particles and run the Higgs mechanism two times. One single Higgsac cannot do the
task, because the scalar particle used for reducing a symmetry should be charged under the
symmetry group we want to reduce. In our case, these symmetry groups are the 
the $U(1)$
factor-groups of \Uo ,\ \Ut\ and \Uth.
Therefore, we begin by first reducing the $U(1)$ sub-groups of \Ut\ and 
\Uth\ to some
residual $U(1)$ whose
corresponding (massless) gauge field will be denoted by $B'_{\mu}$. Subsequently, this symmetry
and the individual \Uo\ will be reduced to the $U(1)$ corresponding to the 
hyper-charge, described
by the gauge field $Y_{\mu}$.

Let us start by choosing the first symmetry-reducing scalar
particle with the transformation properties: 
\be\label{phi1.tr}
\Phi_1(x)\to U_1(x)\Phi_1(x)V_1^{-1}(x)\ , 
\ee 
where $U_1$ stands
for the elements of the $U_3(1)$ sub-group of \Uth\ and $V_1$
stands for the elements of the $U_2(1)$ sub-group of \Ut. We note
that these sub-groups are constructed in the same way as in the
previous section and in Eq.(\ref{phi1.tr}) prodcuts are the usual commutative 
ones. 
The covariant derivative corresponding to this scalar
field is:
\be
D_{\mu}=\partial_{\mu} +{i\over 2}3g_3G^{'0}_{\mu}-{i\over 2}2g_2W^{'0}_{\mu}\ .
\ee
Note that here the covariant derivative only involves the $\theta$-independent 
parts of the corresponding gauge fields. The Lagrangian  for the $\Phi_1$ 
field will acquire the new terms:
\be\label{Phi1}
(D_{\mu}\Phi_1)^{\dagger}(D_{\mu}\Phi_1)+m_1^2\Phi_1^{\dagger}\Phi_1-{f_1\over
4!}(\Phi_1^{\dagger}\Phi_1)^2\ \ \  ({\rm with\ no\ \star-product})\ , 
\ee
which as desired is fully gauge invariant (for more details see Appendix A).
Through the Higgs mechanism, we obtain a mass term for the gauge boson $G^0_{\mu}$:
\be\label{g0_mu}
({3\over 2}g_3G^{'0}_{\mu}-g_2W^{'0}_{\mu})^2\phi_{1}^2 =  N_1^2(G^{0}_{\mu})^2\phi_{1}^2 ,
\ee
where $N_1=\sqrt{g_2^2+({3\over 2}g_3)^2}$ and
$\phi_1=\sqrt{{12 m_1^2\over f_1}}$ is the vacuum expectation value for the
$\Phi_1$ Higgsac field.

The massive gauge boson $G^0_{\mu}$ and the residual massless
$U_1(1)$ field, $B'_{\mu}$, can be defined through a rotation by
the angle $\delta_{23}$, 
\be\label{theta23}
\tan\delta_{23}={2g_2\over 3g_3}\ \ee in the $(W^{'0}_{\mu},
\,G^{'0}_{\mu})$ plane, i. e. \ba\label{rotation23'}
G^{0}_{\mu}&=&\cos\delta_{23}G^{'0}_{\mu}-\sin\delta_{23}W^{'0}_{\mu},\cr
B'_{\mu}&=&
\sin\delta_{23}G^{'0}_{\mu}+\cos\delta_{23}W^{'0}_{\mu}\ , \ea
whose reciprocal is: \ba\label{rotation23}
G^{'0}_{\mu}&=&\cos\delta_{23}G^0_{\mu}+\sin\delta_{23}B'_{\mu},\cr
W^{'0}_{\mu}&=& -\sin\delta_{23}G^0_{\mu}+\cos\delta_{23}B'_{\mu}\ . 
\ea
The remaining $U_1(1)$ group is a particular sub-group of
$U_3(1)\times U_2(1)$ obtained through the mixing process. If we
denote the elements of this $U_1(1)$  group by $s(x)$, the second
scalar field, through which we reduce eventually the symmetry to
that of hyper-charge, should transform as: 
\be\label{phi2.tr}
\Phi_2(x)\to s(x)\Phi_2(x)v^{-1}(x)\ \ \ ({\rm with\ no\ \star-product})  
\ee 
and hence its covariant derivative, which only involves the 
$\theta$-independent parts of gauge fields is given by 
\be
D_{\mu}=\partial_{\mu} +{i\over 2}g_0B'_{\mu}-{i\over 2}g_1B_{\mu}\ ,
\ee
where $g_0=2g_2 3g_3/\sqrt{(2g_2)^2+(3g_3)^2}$ is the coupling constant to the residual
$B'_{\mu}$ field.
Following exactly the same prescription as before for the Higgs mechanism (i.e.
assuming the Lagrangian for the $\Phi_2$ field to be similar to that of $\Phi_1$, given by
(\ref{Phi1})), we shall end up with a new gauge boson, $W^0_{\mu}$, whose mass term in the
Lagrangian will read:
\be\label{W0_mu}
{1\over 4}(g_0B'_{\mu}-g_1B_{\mu})^2\phi_{2}^2 =  N_2^2(W^{0}_{\mu})^2\phi_{2}^2 ,
\ee
where $N_2={1\over 2}\sqrt{g_0^2+g_1^2}$ and $\phi_2$ is the vacuum expectation
value for $\Phi_2$. The massive field, $W^0_{\mu}$, is
related to the fields
$B_{\mu},\ B'_{\mu}$ through a rotation in the $(B_{\mu},\,B'_{\mu})$ plane by the angle
$\delta_{11'}$:
\ba\label{rotation11''}
W^0_{\mu}&=&\cos\delta_{11'}B'_{\mu}-\sin\delta_{11'}B_{\mu},\cr
Y_{\mu}&=& \sin\delta_{11'}B'^0_{\mu}+\cos\delta_{11'}B_{\mu}\ .
\ea
The inverse of this \trans, which relates $W^0$ and  $Y$ (the hyper-photon field) to
$B'$ and $B$, is:
\ba\label{rotation11'}
B'_{\mu}&=&\cos\delta_{11'}W^0_{\mu}+\sin\delta_{11'}Y_{\mu},\cr
B_{\mu}&=& -\sin\delta_{11'}W^0_{\mu}+\cos\delta_{11'}Y_{\mu}\ .
\ea

To summarize, we have reduced the three $U_n(1)$ factors to a
single $U_1(1)$ through two proper Higgsac fields, $\Phi_1$ and
$\Phi_2$ (in principle, in two different energy scales); in the
end, instead of the corresponding three $U(1)$ fields, $G^{'0},\
W^{'0}$ and $B^{0}$, we have introduced two massive gauge bosons,
$G^{0}$ and $W^{0}$ and the (massless) hyper-photon $Y$. The
initial and final $U(1)$ gauge fields are hence related by a
$3\times 3$ rotation matrix $R$: \be\label{R3by3} \left(\matrix {
G^{'0}_\mu \cr
                W^{'0}_\mu \cr
                B_\mu  }\right)= R_{3\times 3}\,
\left(\matrix { G^{0}_\mu \cr
                W^{0}_\mu \cr
                Y_\mu  }\right)\ ,
\ee
where
\be
R_{3\times 3}= R_{23}\ R_{11'}\ ;
\ee
\be
R_{23}=\left(\matrix {\cos\delta_{23}  & \sin\delta_{23}  & 0  \cr
                      -\sin\delta_{23}  & \cos\delta_{23} & 0 \cr
                         0 &  0       & 1 }\right)\ \ , \ \ \ \ \
R_{11'}=\left(\matrix {1 &  0       & 0 \cr
                       0 & \cos\delta_{11'}  & \sin\delta_{11'}   \cr
                       0& -\sin\delta_{11'}  & \cos\delta_{11'}  }\right)\ .
\ee
It is clear from the form of (\ref{R3by3}) that it does not matter in which order we
reduce the $U(1)$ symmetries.

The masses of the massive gauge bosons depend on the $\Phi_1$ and $\Phi_2$ vacuum expectation values:
\be\label{GWmasses}
m_{G^0}=\sqrt{g_2^2+({3\over 2}g_3)^2}\ |\phi_{1}|\ ,\ \
m_{W^0}=\sqrt{({1\over 2}g_1)^2+g_2^2+({3\over 2}g_3)^2}\ |\phi_{2}|\ .
\ee
Then, it is straightforward to rewrite the action (\ref{G123-action}) in terms of the physical
gauge fields: ($G^a_\mu,\ W^i_\mu,\ Y_\mu;\ G^0_\mu,\ W^0_\mu$). We still need to define
the
$Z$ and photon fields out of them, but we will not work it out here and we postpone it to
the next section,  where we discuss the electro-weak symmetry breaking.
However, we would like to comment
that, as we discussed in section 3, the Lagrangian that one will find for the hyper-photon
$Y_\mu$ (upon insertion of (\ref{R3by3}) into (\ref{G123-action})) is 
{\it not} of
the form specific for a pure \Uo\ theory.

After the two Higgsac reductions we end up with the
$NC(SU_c(3)\times SU_L(2)\times U_Y(1))$, where similar to the NC(QED+QCD)
case, this group is in fact a group which in the $\theta\to 0$ limit
recovers the usual SM. However, for non-zero $\theta$ it receives some \nc
corrections. In fact this algebra is the enveloping algebra of the
usual $SU_c(3)\times SU_L(2)\times U_Y(1)$ algebra defined by
insertion of the star products.

\subsection{The matter content}

When coupling the matter fields to the \Uth$\times$\Ut$\times$\Uo\ theory, we have to
keep in
mind that, according to \cite {nogo} (see {\it iii)} in subsection 2.1),
since we have three simple factors in our group,  we can
have only  ${1\over 2}3\times(3+1)=6$ types of charged particles, in the fundamental and/or
anti-fundamental representation
of the group factors. We note that the symmetry-reducing scalar particles (Higgsac
fields) used
in the previous
subsection  which have
to be charged under $U(1)$ {\it factor-groups} of \Uo ,\ \Ut\ and \Uth, are 
not 
included  among these six
types of particles.

Let us now give the gauge transformation properties of these matter fields, together with their
corresponding covariant derivatives.

{\bf 1) Right-handed charged leptons} (in \af\ \rep\ of \Uo). In this group we consider the \rh\
electron, which transforms as
\be
e_R (x)\ \to\ e_R (x)\, v^{-1} (x)\ ,
\ee
and hence the corresponding covariant derivative is
\be\label{coveR}
D^{1}_{\mu}e_R (x)=\partial_{\mu}e_R (x)-{i\over 2}g_1e_R (x)B_{\mu}.
\ee

{\bf 2) Left-handed leptons} (in \fun\ \rep\ of \Ut\ and \af\ \rep\ of \Uo). Here we shall
include the \lh\ electron  and its neutrino, in a doublet:
\be\label{covlL}
\Psi^l_L (x)=\left(\begin{array}{c}\nu (x)\\ e (x)\end{array}\right)_L\ .
\ee
Under the gauge transformations, the doublet transforms as
\be
\Psi^l_L (x)\ \to\ V(x)\, \Psi^l_L (x)\, v^{-1} (x)
\ee
and therefore the corresponding covariant derivative is
\be
D^{1+2}_{\mu}\Psi^l_L (x)=\partial_{\mu}\Psi^l_L (x)+{i\over 2}g_2W^{'0}_{\mu}\Psi^l_L (x)
+{i\over 2}g_2W^{i}_{\mu}\sigma_i\Psi^l_L (x)-{i\over 2}g_1\Psi^l_L (x)B_{\mu}\ .
\ee

{\bf 3,4) Right-handed quarks}. Here, we choose the \rh\ up quark in the \fun\ \rep\ of
\Uo\ and \af\ \rep\ of \Uth\ and the
\rh\ down quark in \af\ \rep\ of \Uth:
\ba\label{Rq.gt}
u_R (x)\ &\to&\ v(x) u_R(x)\, U^{-1} (x)\ ,\cr
d_R (x)\ &\to&\ d_R (x)\, U^{-1} (x)\ ,
\ea
with the covariant derivatives
\be\label{covuR}
D^{1+3}_{\mu}u_R (x)=\partial_{\mu}u_R (x) +{i\over 2}g_1B_{\mu}u_R (x)-{i\over 2}g_3u_R
(x)G^{'0}_{\mu}
-{i\over 2}g_3u_R (x)G^{a}_{\mu}T^a\ ,
\ee
\be\label{covdR}
D^{3}_{\mu}d_R (x)=\partial_{\mu}d_R (x) -{i\over 2}g_3d_R (x)G^{'0}_{\mu}
-{i\over 2}g_3d_R (x)G^{a}_{\mu}T^a\ .
\ee

\vskip .2cm

{\bf 5) Left-handed quarks} - the doublet of \lh\ up and down quarks,
\be
\Psi^q_L (x)=\left(\begin{array}{c} u (x)\\ d(x)\end{array}\right)_L\ ,
\ee
in \fun\ \rep\ of \Ut\ and \af\ \rep\ of \Uth:
\be\label{Lq.gt}
\Psi^q_L (x)\ \to\ V(x)\, \Psi^q_L (x)\, U^{-1} (x)\ ,
\ee
with the covariant derivative:
\ba\label{covqL}
D^{2+3}_{\mu}\Psi^q_L (x)=\partial_{\mu}\Psi^q_L (x)
&+&{i\over 2}g_2W^{'0}_{\mu}\Psi^q_L (x)+{i\over 2}g_2W^{i}_{\mu}\sigma_i\Psi^q_L (x) \cr
&-&{i\over 2}g_3\Psi^q_L (x)G^{'0}_{\mu}-{i\over 2}g_3\Psi^q_L (x)G^{a}_{\mu}T^a\ .
\ea

{\bf 6) Higgs doublet}
\be\label{Higgs}
\Phi (x)=\left(\begin{array}{c}\Phi^+ (x)\\ \Phi^0 (x)\end{array}\right)\ ,
\ee
in \fun\ \rep\ of \Ut,
\be\label{PhiGT}
\Phi(x)\ \to\ V(x)\, \Phi (x)\ ,
\ee
with the covariant derivative:
\be\label{covHiggs}
D^{2}_{\mu}\Phi (x)=\partial_{\mu}\Phi (x)
+{i\over 2}g_2W^{'0}_{\mu}\Phi (x)+{i\over 2}g_2W^{i}_{\mu}\sigma_i\Phi (x)\ .
\ee
We stress that the Higgs field interacts with
other matter and gauge fields directly, i.e. in (\ref{PhiGT}) and 
(\ref{covHiggs}) we should use  $\star$-products and the full gauge fields 
(not only their $\theta$-independent parts, as is the case, for Higgsac 
fields). We would also like to remark that the Higgs doublet fits perfectly in 
this picture and also exhausts the
possible types of charged particles allowed by the no-go theorem \cite{nogo}.

Now, let us show how the \Uo\ symmetry reduction solves the hyper-charge quantization
problem. This fact will
be made obvious by showing that the coupling of all matter fields to the massless gauge field
$Y_{\mu}$ of the residual \Uo\ is realized through the usual hyper-charges of the particles (see
table 1). To this end, we
consider one by one the relevant terms of the Lagrangian
(i.e. $\bar{\Psi}\gamma^{\mu}D_{\mu}\Psi$) for each
type of matter field. In what follows, we  denote the coupling constant to the hyper-charge \Uo\
by $g'$.
The order in which we discuss the different types of fields is the most convenient one:

{\bf i) Right-handed electron}: the coupling to $Y_{\mu}$ can be read off by using
(\ref{rotation11'}):
\ba
\bar{e}_R\gamma^{\mu}D^{1}_{\mu}e_R&=&\bar{e}_R\gamma^{\mu}\partial_{\mu}e_R
-{i\over 2}g_1\bar{e}_R\gamma^{\mu}e_RB_{\mu}\cr
&=&\bar{e}_R\gamma^{\mu}\partial_{\mu}e_R
-{i\over 2}g_1\cos\delta_{11'}\bar{e}_R\gamma^{\mu}e_RY_{\mu}\ +\cdots\ ,
\ea
where the dots contain the coupling to the massive gauge bosons, $G^0_{\mu}$ and $W^0_{\mu}$.
As the coupling term should be proportional to the hyper-charge of $e_R$, i.e. $-2g'$, we
{\it define}
$g'$ as:
\be\label{g'fromg1}
{1\over2}g_1\cos\delta_{11'}=g'\ .
\ee
Also, we note that using the definition of $\delta_{11'}$ mixing angle (\ref{rotation11''}) we
have:
\be\label{tandelta11'}
\cot\delta_{11'}={3g_3\ 2g_2\over g_1\sqrt{(2g_2)^2+(3g_3)^2}}\ ,
\ee
and hence
\be
g'={1\over2}{3g_3\ 2g_2\over\sqrt{(2g_2)^2+(3g_3)^2}}\ \sin\delta_{11'}\ .
\ee

{\bf ii) Right-handed down quark}. In the same way as above, using (\ref{covdR}) and
(\ref{R3by3}), we obtain:
\ba
\bar{d}_R\gamma^{\mu}D^{3}_{\mu}d_R &=&\bar{d}_R\gamma^{\mu}\partial_{\mu}d_R
-{i\over 2}g_3\bar{d}_R\gamma^{\mu}d_R G^{'0}_{\mu}-{i\over 2}g_3\bar{d}_R\gamma^{\mu}d_R
G^{a}_{\mu}T^a \cr
&=&\bar{d}_R\gamma^{\mu}\partial_{\mu}d_R
-{i\over 2}g_3\sin\delta_{23}\bar{d}_R\gamma^{\mu}d_RB'_{\mu}-\cdots \cr
&=&\bar{d}_R\gamma^{\mu}\partial_{\mu}d_R
-{i\over 2}g_3\sin\delta_{23}\sin\delta_{11'}\bar{d}_R\gamma^{\mu}d_R Y_{\mu}-\cdots\ .
\ea
{}From here one can readily find the hyper-charge of the \rh\ down quark,
$Y_{d_R}$:
\be\label{g'fromg3}
g_3\sin\delta_{23}\sin\delta_{11'}=-Y_{d_R}\ .
\ee
However, using (\ref{tandelta11'}) we find that
\be
Y_{d_R}=-{2\over3}g'\ ,
\ee
which is exactly the values given in table 1.
In the following, we show that all the other hyper-charges will also come correctly, using
(\ref{theta23}), (\ref{g'fromg1}) and (\ref{tandelta11'}).

{\bf iii) Right-handed up quark}. Similarly as before
\ba
\bar{u}_R\gamma^{\mu}D^{1+3}_{\mu}u_R &=&\bar{u}_R\gamma^{\mu}\partial_{\mu}u_R
+{i\over 2}g_1\bar{u}_R\gamma^{\mu}B_{\mu}u_R -{i\over 2}g_3\bar{u}_R\gamma^{\mu}u_R
G^{'0}_{\mu}-\cdots\ ,\cr
&=&\bar{u}_R\gamma^{\mu}\partial_{\mu}u_R
+{i\over
2}(g_1\cos\delta_{11'}-g_3\sin\delta_{23}\sin\delta_{11'})\bar{u}_R\gamma^{\mu}Y_{\mu}u_R \cr
&-&{i\over 2}g_3\sin\delta_{23}\sin\delta_{11'}\bar{u}_R\gamma^{\mu}[Y_{\mu},u_R]_{\star} -\cdots\ ,
\ea
from where it emerges that
\be
g_1\cos\delta_{11'}-g_3\sin\delta_{23}\sin\delta_{11'}=Y_{u_{R}}\ ,
\ee
where $Y_{u_{R}}$ is the hyper-charge of the \rh\ up quark. Using (\ref{g'fromg1}) and
(\ref{tandelta11'}), we find:
\be
Y_{u_{R}}={4\over3}g'\ .
\ee

{\bf iv) Left-handed leptons}. For the doublet of \lh\ leptons, we find:
\ba
\bar{\Psi}^l_L\gamma^{\mu}D^{1+2}_{\mu}\Psi^l_L &=&\bar{\Psi}^l_L
\gamma^{\mu}\partial_{\mu}\Psi^l_L
+{i\over 2}g_2\bar{\Psi}^l_L \gamma^{\mu}W^{'0}_{\mu}\Psi^l_L
-{i\over 2}g_1\bar{\Psi}^l_L \gamma^{\mu}\Psi^l_L B_{\mu}+\cdots \cr
&=&\bar{\Psi}^l_L \gamma^{\mu}\partial_{\mu}\Psi^l_L
-{i\over 2}(g_1\cos\delta_{11'}-g_2\cos\delta_{23}\sin\delta_{11'})
\bar{\Psi}^l_L \gamma^{\mu}\Psi^l_L Y_{\mu}\cr
&-&{i\over 2}g_2\cos\delta_{23}\sin\delta_{11'}\bar{\Psi}^l_L\gamma^{\mu}[\Psi^l_L,Y_{\mu}]_{\star}+\cdots\ ,
\ea
from where we read off the condition
\be\label{Yleftlept}
g_1\cos\delta_{11'}-g_2\cos\delta_{23}\sin\delta_{11'}=-Y_{\Psi^l_L}\ .
\ee
Using (\ref{theta23}), (\ref{g'fromg1}) and (\ref{g'fromg3}), from eq. (\ref{Yleftlept}) we obtain:
\be
Y_{\Psi^l_L}=-g'\ .
\ee

{\bf v) Left-handed quarks}. In this case, the relevant coupling term will read:
\ba
\bar{\Psi}^q_L\gamma^{\mu}D^{2+3}_{\mu}\Psi^q_L
&=&\bar{\Psi}^q_L\gamma^{\mu}\partial_{\mu}\Psi^q_L
+{i\over 2}g_2\bar{\Psi}^q_L\gamma^{\mu}W^{'0}_{\mu}\Psi^q_L
-{i\over 2}g_3\bar{\Psi}^q_L\gamma^{\mu}\Psi^q_L G^{'0}_{\mu}+ \cdots\cr
&=&\bar{\Psi}^q_L\gamma^{\mu}\partial_{\mu}\Psi^q_L
+{i\over
2}(g_2\cos\delta_{23}-g_3\sin\delta_{23})\sin\delta_{11'}\bar{\Psi}^q_L\gamma^{\mu}Y_{\mu}\Psi^q_L\cr
&+&{i\over 2}g_3\sin\delta_{23}\sin\delta_{11'}\bar{\Psi}^q_L\gamma^{\mu}[Y_{\mu},\Psi^q_L]_{\star}+\cdots\
,
\ea
which implies that
\be
(g_2\cos\delta_{23}-g_3\sin\delta_{23})\sin\delta_{11'}=Y_{\Psi^q_L}
\ee
and, recalling (\ref{theta23}), (\ref{g'fromg1}) and (\ref{g'fromg3}), the hyper-charge of the
\lh\ quark
doublet in units of $g'$ is found to be
\be
Y_{\Psi^q_L}={1\over3}g'\ .
\ee

{\bf vi) Higgs doublet}. For the last one of the possible charged particles of our model,
\ba
\bar{\Phi}\gamma^{\mu}D^{2}_{\mu}\Phi&=&\bar{\Phi}\gamma^{\mu}\partial_{\mu}\Phi
+{i\over 2}g_2\bar{\Phi}\gamma^{\mu}W^{'0}_{\mu}\Phi+\cdots\cr
&=&\bar{\Phi}\gamma^{\mu}\partial_{\mu}\Phi
+{i\over 2}g_2\cos\delta_{23}\sin\delta_{11'}\bar{\Phi}\gamma^{\mu}Y_{\mu}\Phi+\cdots\ ,
\ea
implying
\be
g_2\cos\delta_{23}\sin\delta_{11'}=Y_{\Phi}
\ee
and eventually, with the help of (\ref{theta23}), (\ref{g'fromg1}) and (\ref{g'fromg3}),
\be
Y_{\Phi}=g'\ .
\ee

Before proceeding with the electro-weak symmetry breaking, let us recount the number of
parameters that we have introduced: There are three different couplings, $g_1,\ g_2$ and $g_3$
which correspond to the
\Uo,\ \Ut\ and \Uth\ factors, respectively. In addition we have introduced two mixing angles,
$\delta_{23}$ and $\delta_{11'}$. However, the physical couplings are $g_2$, the weak coupling,
$g_3$, the strong coupling and $g'={1\over 2}g_1\cos\delta_{11'}$, the hyper-photon coupling.
Also, there are two relations between the couplings and these mixing angles:
\be\label{angles}
\tan\delta_{23}={2g_2\over 3g_3}\ ,\ \ \
\sin\delta_{11'}={g'\over g_2}\sqrt{1+({2g_2\over 3g_3})^2}\ .
\ee
Therefore, both of the mixing angles can be expressed in terms of the physical couplings $g',\
g_2$ and $g_3$.

Here we have chosen a specific order for the symmetry reductions and
the Higgsac fields, namely, first we reduced the $U(1)$'s of
$U_\star(3)\times U_\star(2)$ and then the resulting $U(1)$ with the
extra \Uo\ that we have in our gauge group. We
would like to comment that the choice of any
possible two Higgsac fields as well as
the order of the symmetry reduction(s) do not change the charge
assignments for the quarks and leptons. (Essentially these charges only
depend on the representations of the particles and the fact that we start
with $U_{\star}(3)\times U_{\star}(2)\times U_{\star}(1)$ groups.)

\newsection{The electroweak symmetry breaking}

So far, starting from the \Uott\ gauge theory and reducing two $U(1)$ 
factors, we have arrived at a
theory which can be called $NC(SU_c(3)\times SU_L(2)\times U(1))$. In order to complete
the
formulation of the NCSM, still we should proceed with the usual symmetry breaking through the
Higgs doublet. In fact, by this symmetry breaking, fermions become massive through the Yukawa
terms, which are also allowed in the \nc case.
However, a more important role of this symmetry breaking is to give masses to the $W^i_{\mu},\
i=1,2,3$ fields and also to define the massless photon and massive $Z_{\mu}$ through a
combination of $Y_{\mu}$ and $W^3_{\mu}$.

In this section, first we work out the details of this symmetry breaking in the gauge bosons
sector and then in subsection 5.2, we present the interaction terms of fermions with the
physical gauge bosons, as well as the corresponding Yukawa terms. We also compare these
interaction terms with those of the usual \sm.

For performing the electro-weak symmetry breaking, we  use a doublet of scalar fields of the type
(\ref{Higgs}), charged under
the \Ut\ symmetry group (before the reduction of the $U(1)$ factors
of \Uott). 
Practically, after the $U(1)$ symmetry reduction, this doublet would carry 
hyper-charge and weak
charge. The new terms occurring in the full electro-weak Lagrangian before the symmetry breaking and
due to the
presence of the
doublet of scalar fields are:
\be\label{lagr.Higgs}
{\cal L}_{Higgs}=(D_{\mu}\Phi)^{\dagger}(D_{\mu}\Phi)+\mu^2\Phi^{\dagger}\Phi-{f\over
4!}(\Phi^{\dagger}\Phi)_\star^2+{\cal L}_{Yukawa}\ ,
\ee
where
\ba\label{Yukawa}
{\cal L}_{Yukawa}&=&h_e\bar{e}_R\Phi^{\dagger}\Psi^l_L +h_e^*\bar{\Psi}^l_L \Phi e_R\cr
&+&h_d\bar{d}_R\Phi^{\dagger}\Psi^q_L +h_d^*\bar{\Psi}^q_L \Phi d_R\cr
&+&h_u\bar{u}_R(\Phi^c)^{\dagger}\Psi^q_L +h_u^*\bar{\Psi}^q_L\Phi^c u_R\ 
\ea
and $h_e$, $h_d$ and $h_u$ are the respective Yukawa couplings.
In (\ref{Yukawa}), $\Phi^c$ is the charge conjugated field of $\Phi$, which transforms as:
\be\label{Phicc}
\Phi^c\ \to\ V(x)\, \Phi^c\, v^{-1}.
\ee
Noting (\ref{PhiGT}), the relation (\ref{Phicc}) may seem unusual. However, we recall that
the \nc
particles, besides the usual charge, also carry higher-pole charges and in particular the dipole
charge. Therefore, the charge conjugate of any particle is a particle which is carrying
the opposite of all these higher-pole charges, as well as the charge itself. In fact, one
can check that the charge conjugate of Higgs, $\Phi^c$, should transform as (\ref{Phicc}).

One should also note that it is not possible to construct the Yukawa terms in the Lagrangian
corresponding to the $U(1)$ symmetry-reducing Higgsac fields $\Phi_1$ and 
$\Phi_2$, because no
gauge invariant combination of them with the fermionic fields
could exist.

The potential of (\ref{lagr.Higgs}) has a minimum at $(\Phi^{\dagger}\Phi)_0={12\mu^2\over
f}\equiv\phi_0^2$ 
and we can choose the vacuum expectation value for the scalar field to be\footnote{ We note
that, since this minimum is  $x$-independent, one can drop the \sp s and hence the minimum-energy 
solution is the same as in the \com\ case.} 
\be
\Phi_0=\left(\begin{array}{c}0\\ \phi_0\end{array}\right)\ .
\ee 

\subsection{Symmetry breaking in the gauge bosons sector}

Now we  discuss the details of the electro-weak symmetry breaking and its implications on the gauge
bosons sector. To this end, we write the full covariant derivative of the Higgs field, which is the main
ingredient of the mass-generating term. Having in view (\ref{rotation23}),
(\ref{rotation11'}) and (\ref{covHiggs}), we obtain:
\ba
D^{2}_{\mu}\Phi (x)&=&\partial_{\mu}\Phi (x) 
+{i\over 2}[g'Y_{\mu}+\sin\delta_{23}(-g_2G_{\mu}^0+{3\over2}g_3\cos\delta_{11'}W^{0}_{\mu})]\Phi
(x)\cr
&+&{i\over 2}g_2W^{i}_{\mu}\sigma_i\Phi (x)\ .
\ea
Hence, the mass term emerging from here is of the form of
\be\label{masslagr}
L_M=\left\{\big[(g'Y_{\mu}-g_2W^3_{\mu})+\sin\delta_{23}(-g_2G_{\mu}^0
+{3\over2}g_3\cos\delta_{11'}W^{0}_{\mu})\big]^2
+2g_2^2W^+_{\mu}W^-_{\mu}\right\}|\phi_0|^2\ ,
\ee
where $W^{\pm}={1\over{\sqrt2}}(W^1\pm iW^2)$.

Following the usual Higgs mechanism, let us identify the first term in the brackets in
(\ref{masslagr}) as being proportional to
$Z_{\mu}^0$; more explicitly
\be\label{weakrot}
Z_{\mu}^0=-\sin\theta^0_W Y_{\mu}+\cos\theta^0_W W_{\mu}^3\ ,
\ee
\be\label{Photon}
A_{\mu}=\cos\theta^0_W Y_{\mu}+\sin\theta^0_W W_{\mu}^3\ ,
\ee
where the weak mixing angle $\theta^0_W$ is defined as in the usual \sm:
\be\label{theta0}
\tan\theta^0_W={g'\over g_2}\ .
\ee 
With these notations, we can rewrite (\ref{masslagr}) as:
\be\label{newmasslagr}
L_M=\left\{\big[gZ^0_{\mu}+\sin\delta_{23}(-g_2G_{\mu}^0
+{3\over2}g_3\cos\delta_{11'}W^{0}_{\mu})\big]^2
+2g_2^2W^+_{\mu}W^-_{\mu}\right\}|\phi_0|^2\ ,
\ee
where 
\be\label{g}
g=\sqrt{g^{'2}+g_2^2}\ .
\ee

In (\ref{newmasslagr}), the mass term for the $W^{\pm}$ bosons is clearly singled out:
\be
m_{W^{\pm}}=g\cos\theta^0_W|\phi_0|\ .
\ee
As we see from (\ref{newmasslagr}), $Z^0_{\mu}$ is not the real physical $Z$-boson. The
physical $Z$-particle, which diagonalizes the above mass-Lagrangian is mixed with the other
two massive gauge bosons,  $W_{\mu}^0$ and $G_{\mu}^0$. As a result, we have a correction to
the mass of the physical $Z$-particle, compared to the usual \sm. 
However, still the massless gauge boson, the photon, is given by (\ref{Photon}).

In order to compute this correction, we have to take into account also the mass terms of $W_{\mu}^0$ 
and $G_{\mu}^0$ obtained during the $U(1)$ symmetry breaking and to 
diagonalize the obtained mass matrix.
Recalling (\ref{GWmasses}) and (\ref{g}), we can write the mass term for $Z_{\mu}^0$,
 $W_{\mu}^0$ and $G_{\mu}^0$:
\ba
L_{ZWG}&=&{g^2\over2}\Big\{\big[Z^0_{\mu}+\cos\theta^0_W(\sin\delta_{23}G_{\mu}^0
-\cos\delta_{11'}\cos\delta_{23}W^{0}_{\mu})\big]^2|\phi_0|^2\cr
&+&\big(\frac{\cos^2\theta^0_W}{\sin^2\delta_{23}}\phi_1^2\big)G_{\mu}^{02}
+\big(\frac{\cos^2\theta^0_W\cos^2\delta_{23}}{\cos^2\delta_{11'}}\phi_2^2\big)W_{\mu}^{02}
\Big\}\cr
&\equiv&{g^2\over2}\phi_0^2\, {\bf X^tMX}\ ,
\ea
where
\be\label{Massmatrix}
{\bf X}=\left(\matrix { Z^{0}_\mu \cr
                G^{0}_\mu \cr
                W^{0}_\mu  }\right)\ \ , \ \ \ \ \ 
{\bf M}=\left(\matrix {1 &  a       & -b \cr 
                       a & a^2+d^2  & -ab   \cr
                       -b& -ab  & f^2+b^2  }\right)\ , 
\ee  
with
\ba
a=\cos\theta^0_W\sin\delta_{23}\ &,& \ \ \
b=\cos\theta^0_W\cos\delta_{11'}\cos\delta_{23}\ ,\cr
d=\frac{\cos\theta^0_W}{\sin\delta_{23}}\frac{\phi_1}{|\phi_0|}\ &,& \ \ \
f=\frac{\cos\theta^0_W\cos\delta_{23}}{\cos\delta_{11'}}\frac{\phi_2}{|\phi_0|}\ .
\ea

Since the physical $Z$-field, $Z_{\mu}$, and $Z^0_{\mu}$  should almost be equivalent,
we expect the $a$ and $b$ factors  of (\ref{Massmatrix}) to be small (compared to
$d$ and $f$).
Physically, this is equivalent to assuming that 
$$
{m_Z\over m_{G^0}},\ {m_Z\over m_{W^0}} \ll 1\ .
$$
Then, diagonalizing (\ref{Massmatrix}), the  mass for the physical $Z$-particle, up to the
second order in ${m_Z\over m_{G^0}},\ {m_Z\over m_{W^0}}$ is obtained to be
\be\label{Zmass}
m^2_Z=g^2  |\phi_0|^2
\left[1-\sin^4\delta_{23}(\frac{\phi_0}{\phi_1})^2-\cos^4\delta_{11'}(\frac{\phi_0}{\phi_2})^2 \right]\ ,
\ee
and therefore, 
\be\label{massratio}
\frac{m^2_W}{m^2_Z}=\cos^2\theta^0_W\left\{1+\cos^2\theta^0_W[(\frac{m_Z}{m_{G^0}})^2\sin^2\delta_{23}
+(\frac{m_Z}{m_{W^0}})^2\cos^2\delta_{23}\cos^2\delta_{11'}]\right\}\ .
\ee
We also note that, using (\ref{angles}) and
(\ref{theta0}), we have: $\cos^2\delta_{23}\cos^2\delta_{11'}=\cos^2\delta_{23}-\tan^2\theta^0_W$.

Having identified the physical gauge fields: the massless gluons, $G^a_{\mu},\ a=1,2,\cdots,8$; the photon,
$A_{\mu}$; the massive gauge bosons $W^\pm_\mu, Z_\mu, W^0_\mu$ and $G^0_\mu$, one can rewrite the action 
(\ref{G123-action}) in terms of  these fields and the corresponding couplings. Although we do not write the
latter down
here explicitly, we would like to comment that in the \nc case we have three and four -photon
interaction vertices, which are {\it not} the vertices arising from a pure \Uo\
theory.
Besides the differences in the photon-photon vertices, there exist also photon-$Z$ interaction terms which have no counter-part in
the \sm.

\subsection{Symmetry breaking in the fermionic sector}

In order to pick up the fermionic interaction terms after the electro-weak symmetry
breaking, we shall
explicitly write down the relevant interaction terms of the \Uott\ Lagrangian,
separately for the
leptonic and quark sectors.

For the leptonic sector, using (\ref{coveR}) and (\ref{covlL}):
\ba\label{lagr.ferm}
{\cal L}_{leptons}&=&-{i\over 2}g_1\bar{e}_R\gamma^{\mu}e_RB_{\mu}
+{i\over 2}g_2\bar{\Psi}^l_L \gamma^{\mu}W^{i}_{\mu}\sigma_i\Psi^l_L \cr
&+&{i\over 2}\bar{\Psi}^l_L \gamma^{\mu}(g_2W^{'0}_{\mu}-g_1B_{\mu})\Psi^l_L
-{i\over 2}g_1\bar{\Psi}^l_L \gamma^{\mu}[\Psi^l_L,B_{\mu}]_{\star}
\ea
and for the quark sector, recalling (\ref{covuR}), (\ref{covdR}) and (\ref{covqL}):
\ba\label{lagr.quark}
{\cal L}_{quarks}&=&{i\over 2}g_1\bar{u}_R\gamma^{\mu}B_{\mu}u_R
-{i\over 2}g_3\bar{u}_R\gamma^{\mu}u_RG^{'0}_{\mu}
-{i\over 2}g_3\bar{u}_R\gamma^{\mu}u_RG^{a}_{\mu}T^a\cr
&-&{i\over 2}g_3\bar{d}_R\gamma^{\mu}d_RG^{'0}_{\mu}-{i\over
2}g_3\bar{d}_R\gamma^{\mu}d_RG^{a}_{\mu}T^a\cr
&+&{i\over 2}g_2\bar{\Psi}^q_L\gamma^{\mu}W^{'0}_{\mu}\Psi^q_L 
+{i\over 2}g_2\bar{\Psi}^q_L\gamma^{\mu}W^{i}_{\mu}\sigma_i\Psi^q_L \cr
&-&{i\over 2}g_3\bar{\Psi}^q_L\gamma^{\mu}\Psi^q_LG^{'0}_{\mu}
-{i\over 2}g_3\bar{\Psi}^q_L\gamma^{\mu}\Psi^q_LG^{a}_{\mu}T^a\ .
\ea        
After the reduction of the $U(1)$ 
factors and the electro-weak symmetry breaking, from (\ref{lagr.ferm}) and 
(\ref{lagr.quark}) we obtain the following interaction terms:
\vskip 1cm

{\bf Leptonic sector}
\vskip .5cm
{\it Electron-photon interaction vertex} 
comes in a form analogous to that of the usual \sm:
\be\label{e-gamma}
{\cal L}_{\Psi_e-\gamma}= -i e\bar{\Psi}_e\gamma^{\mu}\Psi_eA_{\mu}\ ,
\ee
with $e$ being the coupling. Following the symmetry breaking procedure, we obtain  
\be\label{el.charge}
e=g_2\sin\theta^0_W={1\over 2}g\ \sin 2\theta^0_W\ ,
\ee  
using (\ref{theta0}) and (\ref{g}). From this form of the interaction term, it is clear that
the electron is in the
\af\ \rep\ of the residual \Uo\ group, described by the massless gauge field $A_{\mu}$ and 
corresponding to NCQED.
\vskip .5cm
{\it Electron-$Z_{\mu}$ vertex}:

\ba\label{e-Z}
{\cal L}_{\Psi_e-Z_{\mu}^0}&=&ig[(-{1\over
2}+\sin^2\theta_W^0)\bar{\Psi}_{e_L}\gamma^{\mu}Z^0_{\mu}\Psi_{e_L}]
+ig\sin^2\theta_W^0\bar{\Psi}_{e_R}\gamma^{\mu}Z^0_{\mu}\Psi_{e_R}\cr
&+&ig\sin^2\theta_W^0\bar{\Psi}_{e}\gamma^{\mu}[\Psi_{e},Z^0_{\mu}]_{\star}\ ,
\ea
where the first two terms are of the same form as in the usual \sm. However, still one should keep in
mind that:
\newline
1) actually what appears in the interaction terms (\ref{e-Z}) is $Z^0_\mu$ and not the physical
$Z$-particle. Hence, these terms also generate extra interaction terms between electron and $G^0$ and
$W^0$ massive gauge bosons.
\newline
2) Still one should be careful with the order of the fields, due to the \sp. 

In particular, we note the Moyal bracket term; indicating that the \nc electron besides the usual
$Z$-charge also couples to the derivatives of $Z-\mu$. I
n the first order in $\theta_{\mu\nu}$,
basically this is a weak-dipole-Z interaction. 
\vskip .5cm
{\it Electron-neutrino-$W_{\mu}^{\pm}$ interaction term}:
\be\label{e-n-W}
{\cal L}_{\Psi_e-\nu-W_{\mu}^{\pm}}=i\sqrt{2}g_2(\bar{\nu}\gamma^{\mu}W^+_{\mu}\Psi_{e_L}
+\bar{\Psi}_{e_L}\gamma^{\mu}W^-_{\mu}\nu)\ ,
\ee
which apart from the \sp s between the fields, is the same as that of the usual \sm.
\vskip .5cm
{\it Neutrino-photon interaction}:
\be\label{n-gamma}
{\cal L}_{\nu-\gamma}=-ie\bar{\nu}\gamma^{\mu}[\nu,A_{\mu}]_{\star}\ ,
\ee
is a completely new interaction, realized through the neutrino dipole moment.
We will elaborate more on this interaction term and its physical consequences in the next section.
\vskip .5cm
{\it Neutrino-$Z_{\mu}$ interaction}:
\ba
{\cal L}_{\nu-Z_{\mu}^0}={i\over 2}g\bar{\nu}\gamma^{\mu}Z_{\mu}^0\nu
+ig\sin^2\theta_W^0\bar{\nu}\gamma^{\mu}[\nu,Z^0_{\mu}]_{\star}\ ,
\ea
where the first term is of the same form as in the \sm. However, the second term is a 
result of the fact that the \nc neutrino also carries $Z$-dipole moment.
\vskip 1cm
{\bf Quark sector}        
\vskip .5cm
{\it Up quark-photon interaction}:
\be\label{up-photon}
{\cal L}_{u-\gamma}={2i\over 3}e\bar{u}\gamma^{\mu}A_{\mu}u
-{i\over 3}e\bar{u}\gamma^{\mu}[u,A_{\mu}]_{\star}.
\ee
As we see the up quark besides simple insertion of the \sp\ also involves another Moyal bracket
term. This extra term which is basically coming from the fact that the up quark is non-singlet
under two
group factors (\ref{Rq.gt}) and (\ref{Lq.gt}), has an interesting consequence: the electric dipole moment
of up quark is twice more than what expected from naive NCQED. To see this let us expand
(\ref{up-photon}) in powers of $\theta_{\mu\nu}$. Up to the first order we have
\be
{\cal L}_{u-\gamma}={2i\over 3}e\bar{u}\gamma^{\mu}A_{\mu}u -{2\over
3}e\bar{u}\gamma^{\mu}\left(\theta_{\alpha\beta}\partial_{\alpha}A_\mu \partial_{\beta} u\right)
+{\cal O}(\theta^2)\ . 
\ee
Recalling the arguments of \cite{Lamb}, one expects to find ${1\over 3}$
for the coefficient of the second term, 
while what we obtain is
${2\over 3}$.
\vskip .5cm
{\it Down quark-photon interaction}:
\be
{\cal L}_{d-\gamma}=-{i\over 3}e\bar{d}\gamma^{\mu}dA_{\mu}\ .
\ee
This is exactly what one expects from a naive extension of QED to NCQED, by insertion of \sp s.
\vskip .5cm 
{\it Up quark-$Z_{\mu}$ interaction}
\ba
{\cal L}_{u-Z_{\mu}^0}&=&ig({1\over 2}-{2\over
3}\sin^2\theta_W^0)\bar{u}_L\gamma^{\mu}Z^0_{\mu}u_L
-ig{2\over 3}\sin^2\theta_W^0\bar{u}_R\gamma^{\mu}Z^0_{\mu}u_R\cr
&+&ig{1\over 3}\sin^2\theta_W^0\bar{u}\gamma^{\mu}[u,Z^0_{\mu}]_{\star}\ .
\ea
Up to the difference in the \sp s (not written explicitly according to our convention), 
the first two terms are the same as in usual \sm, while the third term is again showing
the weak-higher-pole moments of the \nc up quark.
\vskip .5cm
{\it Down quark-$Z_{\mu}$ vertex}

\ba
{\cal L}_{d-Z_{\mu}^0}&=&ig(-{1\over 2}+{1\over
3}\sin^2\theta_W^0)\bar{d}_L\gamma^{\mu}Z^0_{\mu}d_L
+ig{1\over 3}\sin^2\theta_W^0\bar{d}_R\gamma^{\mu}Z^0_{\mu}d_R\cr
&+&ig{1\over 3}\sin^2\theta_W^0\bar{d}\gamma^{\mu}[d,Z^0_{\mu}]_{\star}\ .
\ea
\vskip .5cm

{\it Up quark-down quark-$W_{\mu}^{\pm}$ interaction}:
\be
{\cal L}_{u-d-W_{\mu}^{\pm}}={i\over\sqrt{2}}g_2(\bar{u}_L\gamma^{\mu}W^+_{\mu}d_L
+\bar{d}_L\gamma^{\mu}W^-_{\mu}u_L)\ .
\ee

\newsection{Some specific features of NCSM}

In the previous section,  we worked out in detail the 
fermions-gauge bosons interaction terms in the NCSM. 
In general, one can classify the new ingredients of the NCSM in two sets:

First are those coming from the group theoretical structure of the model and do not depend on
the
\ncy\ parameter explicitly. This set is mainly a consequence of having two extra massive gauge
bosons, $G^0_{\mu}$
and $W^0_{\mu}$. Although we did not present it, almost all the fermions  interact with the new
massive gauge bosons, $G^0_{\mu}$ and $W^0_{\mu}$. Such interaction terms effectively will give
rise to Fermi's four-fermion interaction, where its coupling (up to some numeric factors) is
$G_F({m_Z\over m_{W^0}})^2$. Another important effect of these new massive gauge bosons is the
correction to the physical $Z$-particle, and in particular to its mass. We will discuss this in details
in the subsection 6.2 and in this way we impose some lower bounds on the masses of these new
massive gauge bosons. 

The second class of new features in the NCSM are the interaction terms coming from the \sp\ and
(at least at the classical level) in the \com\ limit, i.e. $\theta\to 0$, they vanish explicitly.
In other words, all the particles, besides the usual charge, up to the first order in
$\theta_{\mu\nu}$, also carry dipole charge which is proportional to the \ncy\
parameter \cite{{Lamb},{Ren}}. From these new interaction terms here we discuss that of 
neutrino-photon coupling and from there we obtain a lower bound on the \ncy\ scale.

\subsection{Neutrino dipole moment}

As we have explicitly shown in the previous section, and in particular in (\ref{n-gamma}), neutrino in the NCSM
undergoes a new type of interaction: the neutrino-photon vertex. Unlike all the other photon-fermion
interactions in the NCSM, this vertex is a {\it chiral} one, i.e. the only existing neutrino, the
left-handed $\nu$, appears in this interaction term. More precisely, in the \nc case, we do
not need necessarily a right-handed neutrino to have a coupling to the electro-magnetic field and therefore the
neutrino, without being massive, can carry dipole charges.

On the other hand, there are very strong (astro-physical) bounds on the neutrino-photon interactions and 
especially the neutrino dipole moment \cite{Data}. In fact, these bounds 
can be translated to a lower bound 
on the \ncy\ scale, $\Lambda_{NC}$, defined as:
\be
\theta_{\mu\nu}={1\over \Lambda^2_{NC}} \epsilon_{\mu\nu}\ ,
\ee
where $\epsilon_{\mu\nu}$ is  a dimensionless anti-symmetric parameter, whose elements
are of the order of one. 

It is well-known that neutrino has a considerable effect in the stellar cooling process. However,
according to
the \sm, they only participate in the weak interactions through massive $W^\pm$ and $Z$. In this way any
direct photon-neutrino interaction such as what we have here, can speed up the cooling process, which in
turn will change the whole star evolution. To avoid drastic changes in this respect (which have
not been observed) the strength of neutrino-photon interaction should be smaller compared to that of $Z$. 
To materialize the above argument, let us expand (\ref{n-gamma}) up to the first order in
$\theta_{\mu\nu}$:
\be
{\cal L}_{\nu-\gamma}=-ie\bar{\nu}\gamma^{\mu}[\nu,A_{\mu}]_{\star}=-e\ 
\bar{\nu}\gamma^{\mu}\left(\theta_{\alpha\beta}\partial_{\alpha}A_\mu \partial_{\beta}\nu\right)
+{\cal O}(\theta^2)\ .
\ee
As we see, in the above interaction the derivative of neutrino appears (as well as that of the
photon field $A_{\mu}$). Then, one can read off the effective neutrino \nc dipole moment:
\be
d_{\nu}= e{1\over \Lambda^2_{NC}}\ E_{\nu}\ , 
\ee
where $E_{\nu}$ is the energy of the neutrino. For the case at hand, the solar neutrino
problem, 
$E_\nu \simeq 10 MeV$ and the corresponding bound on the magnitude of dipole moment is
\cite{neutrino}
\be
d_\nu \lesssim 0.1\times 10^{-10}\ \mu_{B}\ ,
\ee
where $\mu_B={e\hbar\over 2m_e c}$ is the Bohr magneton 
\footnote{In fact this bound is coming from the consideration of Red Giant cooling
process. 
There are some weaker and also stronger bounds on the neutrino dipole moment coming from
some other sources. Since in our model we do not have right-handed neutrinos we cannot
use the stronger bound of $10^{-13}\mu_B$.}. 
Therefore, one can readily
obtain the lower bound on the \ncy\ scale
\be
\Lambda_{NC}\gtrsim  10^3 \ GeV\ .
\ee
Of course, this bound is based on a rough estimate and a more detailed calculation and survey can
improve this bound. Also we note that this bound is of the same order as the previous bounds coming from
the Lamb shift \cite{Lamb} and the Lorentz-violation considerations 
\cite{{Harvey}}.

\subsection{Corrections to the weak-mixing angle}

As we have discussed previously in section five, the physical $Z$-particle, which is an
eigen-state 
of the mass matrix after the electro-weak symmetry breaking, besides the $W^3_{\mu}$, the
 hyper-photon $Y_{\mu}$ now receives a contribution from the other two new massive gauge
bosons, $G^0_{\mu}$ and $W^0_{\mu}$, while the photon field is only made out of $W^3_{\mu}$
and $Y_{\mu}$, in such a way that at the end $Z_{\mu}$ and the photon field $A_{\mu}$ are 
ortho-normal states.
However, as we have explicitly shown, these contributions are suppressed by the 
$({m_Z\over m_{W^0}})^2$ ratio, eq. (\ref{Zmass}). On the other hand, the $W^\pm$ gauge
bosons remain
the same as in the usual \sm, $W^\pm_{\mu}={1\over \sqrt 2}(W^1_{\mu}\pm iW^2_{\mu})$.
Therefore, the ${m_Z\over m_{W}}$ ratio now receives a correction, as indicated in
(\ref{massratio}).
We remind that the weak-mixing angle $\theta^0_W$, is still defined through the ratio of
hyper-photon coupling and the weak coupling: ${g'\over g_2}=\tan\theta^0_W$.

In the usual \sm, although the parameter
$$
\rho=({m_Z\over m_{W}})^2\cos^2\theta^0_W\ 
$$
at classical (tree) level is equal to one, it receives quantum (loop) corrections, see e.g.
\cite{Altarel1}. In fact, one of the precision tests of the \sm\ is to evaluate these corrections to
$\rho$ and compare them to the corresponding experimental data \cite{{Altarel1},{Altarel2}}.
Here we use the conventions and notations of \cite{Altarel1} to parameterize these corrections:
\ba
\left. ({m_Z\over m_{W}})^2=({m_Z\over m_{W}})^2\right|_{B}\ (1+1.43\epsilon_1-1.00
\epsilon_2-0.86
\epsilon_3)\ ,
\ea
where the $\epsilon_i$ show the "large" asymptotic contributions, up to the leading linearized
approximation and
$$
\left. ({m_Z\over m_{W}})^2\right|_{B}=0.768905\ 
$$
is the $Z$ and $W$ mass ratio in the Born approximation.
With the latest data used in \cite{Altarel2}, the {\it predicted} values of $\epsilon$
variables in the usual \sm, which do depend on Higgs and top quark masses,
are given in table 2. 
\begin{table}[ht]\label{tab:epsilon}
\vspace{0.3cm}
\begin{center}
\begin{tabular}{|c|c|c|c|}
\hline
$\epsilon_i\ \times 10^{+3}$ & $m_t=174.3-5.1$ & $m_t=174.3$ & $m_t=174.3+5.1$ \\
\hline
$\epsilon_1\ $ & $5.1$   & $5.6$  & $6.0$ \\
$\epsilon_2\ $ & $-7.2$  & $-7.4$ & $-7.6$ \\
$\epsilon_3\ $ & $5.4$   & $5.4$  & $5.3$   \\
\hline
\end{tabular}
\end{center}
\caption{\sm\ predictions for $\epsilon$ variables, at $m_H=113 GeV$}
\end{table}
However, the observed values of $\epsilon_i$'s obtained from all combined hadronic, leptonic
and Higgs measurements are:
\ba\label{exper}
\epsilon_1 &=& (5.4\pm 1.0) \times 10^{-3}\  ,\cr   
\epsilon_2 &=& (-9.7\pm 1.2) \times 10^{-3}\ ,\cr   
\epsilon_3 &=& (5.4\pm 0.9) \times 10^{-3}\ .
\ea
Comparing the  \sm\ model results and the observed values (\ref{exper}), the \nc corrections
should be smaller than the difference between these two values. More explicitly, 
\be
\cos^2\theta^0_W\Big[(\frac{m_Z}{m_{G^0}})^2\sin^2\delta_{23}
+(\frac{m_Z}{m_{W^0}})^2\cos^2\delta_{23}\cos^2\delta_{11'}\Big]\lesssim (2.014\pm
3.404) \times
10^{-3}\ .
\ee  
On the other hand, 
\ba
\tan\delta_{23}=\left.{2\over 3}\sqrt{{\alpha_{QED}\over \alpha_s}{1\over
\sin^2\theta^0_W}}\ \ \right|_{m_Z} 
= 0.354\ ,
\ea
where in the above we have used the data given in \cite{Data} \footnote{Using the relations 
defining $\delta_{11'}$ we find that: $\ \sin^2\delta_{11'}={\tan^2\theta^0_W\over
\cos^2\delta_{23}}=0.3383$.}.
Now, if we assume that $m_{G^0}\simeq m_{W^0}$, we can find a lower bound on $m_{G^0}$:
\be
m_{G^0} \gtrsim  2.5\times 10 \ m_Z\ .
\ee

\newsection{Outlook}

In this work we have constructed the \nc version of the \sm\
(NCSM). Mainly, the present article is devoted to presenting the
formulation in which the obstacle against such a \nc version of
\sm\ has been overcome. We have classified these problems and
obstacles in three categories; however, the most important one was
the charge quantization problem. We have discussed how this
problem can be resolved, while respecting the no-go theorem
stating that matter fields cannot carry more than two kinds of
charges \cite{nogo}. In fact, as we have shown, {\it only} the
matter content as in the usual \sm\ (including Higgs) is allowed
in a \nc extension. Our recipe to remove these problems is based
on the reduction of the extra $U(1)$ symmetries through the Higgs
mechanism (and Higgsac fields), in which the residual massless
$U(1)$ field becomes a linear combination of the original $U(1)$
fields, (\ref{R3by3}). The detailed discussion of the results in
the $0$-order in $\theta$, compatible with the observed values for
(hyper)charges, is given in the Appendices. We postpone the
complete solution (to all orders in $\theta$) of the $U(1)$
sub-groups reduction to a future work.

Actually, here we have just introduced the NCSM at {\it classical level}
and mainly in the {\it leading order in $\theta$} and not explored all the
possible new features of the NCSM. These are open
questions to be studied in future works.
However, among the new features we have briefly discussed the neutrino
dipole moment which is a natural out-come of our model. This dipole moment
interaction imposes a lower bound on the \ncy\ scale:
\be
\Lambda_{NC}\gtrsim 10^{3} GeV\ .
\ee

We have discussed that there are corrections to the ${m^2_W\over m^2_Z}$ ratio, which do depend on
the masses of the extra gauge bosons. Using the experimental bounds \cite{{Altarel1},{Altarel2}},
we have found the lower
bound on the masses of these gauge bosons:
\be
m_{W^0},\ m_{G^0} \gtrsim 25\ m_Z\ .
\ee

As we see, the bounds on all the three dimensionful parameters of our theory, $\Lambda_{NC},
m_{W^0}$ and $m_{G^0}$ are of the order of $1-10\, TeV$.

The other direct consequence of our model is the inherent CP violation, which is in both the
leptonic (including neutrinos) and quark sectors and is controlled by the \ncy\ parameter,
$\theta_{\mu\nu}$.

On the NCSM, there are  a few other remarks in order:
\newline
1) Most of our arguments for constructing the model in sections 3, 4 and 5, do not depend
on the details of the \sp\ we have used and only having a \nc (but associative) product would
lead to the same conclusion.
\newline
2) Anomaly cancellation:

It is well known that an important theoretical consistency check
for the usual \sm\ (as a chiral gauge theory) is the cancellation
of the triangle anomaly. In fact, this anomaly cancellation is a
consequence of the details of matter content and corresponding
charges. In the \nc case, the anomaly calculations have been
already done in \cite{Anomaly}. According to these works a \nc
gauge theory, in order to be anomaly free, should be vector-like.
Hence, a \nc version of \sm\ is incurably sick. However, along
with the arguments of \cite{phantom}, the mixed anomalies (those
which are of the form of $U_\star(n)-U_\star(n)-U_\star(m),\ m\neq
n$ and also $U_\star(1)-U_\star(2)-U_\star(3)$) are not present.
Furthermore, our theory in the $U_\star(3)$ sector is vector-like.
Although it is not clear how, we believe that the other two
anomalous diagrams ($(U_\star(1))^3$ and $(U_\star(2))^3$) can be
removed. One possible way, among others, as discussed in
\cite{phantom} can be making the supersymmetric version of NCSM.
We hope that  using the effective $NCSU(n)$ groups defined here we
can solve the anomaly problem. We postpone a full analysis of the
anomaly problem to future works.
\newline
3) Quarks mixings:

Although we have not considered them here, usual quarks mixings are also possible in the NCSM.
If we only consider the usual unitary CKM mixing matrix (whose entries are constant and not
space-time functions), the \nc effects will appear only at the loop level. (The \ncy\ appears as
some overall phases in the amplitudes and hence in the probability and cross sections it will go away.)
\newline
4) Neutrino mass and mixing:

In our model, neutrinos are massless, however, we can add masses and mixings. According to the
no-go theorem, since we have exhausted all six possibilities for particles carrying any kind of
charge, we cannot have a right-handed neutrino which carries a charge.
Hence, the \rh\ neutrino could only be a sterile neutrino, i.e. a singlet under all the \Uo, \Ut\ and
\Uth\ factors and could appear only through the mixing with active neutrinos, or it could be a dipole of
one of the group factors, among which the most plausible is the \Uo\ factor, i.e. $\nu_R\to v\nu_R
v^{-1}$.

Finally, as an immediate check for our model, one should examine the running of the \nc photon
coupling, and as we have discussed, there is a reasonable hope to resolve the negative $\beta$-function
problem of NCQED mentioned in \cite{Haya}.

{\it Note added}: After this paper was submitted to the hep-archive
(hep-th/0107055), another very interesting work with the same main
subject, by X. Calmet, B. Jurco, P. Schupp, J. Wess and M. Wohlgennant
has appeared \cite{Wess}. In this
work and also its follow-ups \cite{SWmap}, the construction of the NCSM is 
based on the Seiberg-Witten map and it essentially differs from our 
approach in the fact that the
internal symmetries are considered at the level of the algebra, while in
our case they are considered at the gauge-group level. It is indeed very
interesting to find and account for the different effects emerging from
these two different approaches.

\vskip 1cm

{\large{\bf Acknowledgments:}}

M.C. and A.T. are indebted to A. Kobakhidze and C. Montonen for
enlightening remarks. M.M. Sh-J. would like to thank
Y. Farzan, A. Smirnov and J. Ellis for fruitful discussions.
The financial support of the Academy of Finland under the Project Nos.
163394 and 54023
is greatly acknowledged. The work of P.P. was partially supported by VEGA
project 1/7069/20.

\vskip 1cm

\appendix

{\LARGE{\bf Appendices}}
\vskip .5cm

\section{Normal sub-groups of $U_{\star}(n)$}
\setcounter{equation}{0}
Let ${\cal A}_{\star}$ be the Lie algebra of function on the
Moyal plane generated by commutators. For any $k=0,1,\dots$, we
define recursively:
\be\label{6}
{\cal A}^{\star}_{k+1}\ =\ \{{\cal
A}^{\star}_k,{\cal A}_{\star}\}_{MB}\ ,\ {\cal A}^{\star}_0\ \equiv\ {\cal
A}_{\star}\ .
\ee

Any $f(x)\in{\cal A}_{\star}$ is a power series in
$\theta_{\mu\nu}$, the set ${\cal A}^{\star}_k$ is formed by power
series {\it starting} with the $k$-th power. The set of
sub-algebras ${\cal A}^{\star}_k$, $k=0,1,\dots$, form a {\it filtration}
of the Moyal plane: ${\cal A}^k_{\star} {\cal A}^m_{\star}\subset{\cal
A}^{k+m}_{\star}$.

The gauge algebra ${\cal B}=u_{\star}(n)$ is defined as the
following set of matrix functions on ${\cal A}_{\star}$:
\be\label{7} 
\epsilon(x)\ =\ \epsilon_0(x){\bf
1}_n+\epsilon_a(x)T^a\ \equiv\ \epsilon_A(x)T^A\ ,
\ee 
where
$T^0={\bf 1}_n$ is $n\times n$ unit matrix and $T^a$, $a=1,\dots
,n^2-1$, are $n\times n$ Gell-Mann matrices satisfying the relations:
$T^AT^B= \delta^{AB}{\bf 1}_n+(d^{AB}_C+if^{AB}_C)T^C$;
$\epsilon_a(x)= \epsilon^\dagger_a(x)$ are hermitian functions
from ${\cal A}_{\star}={\cal A}^{\star}_0$.

Let us consider now the commutator algebra ${\cal B}'=[{\cal
B},{\cal B}]$.  This is an ideal formed by elements of
$u_{\star}(n)$ with $\epsilon_0(x)\in{\cal A}^{\star}_1$ and
$\epsilon_A(x)\in{\cal A}_{\star}$. Further, ${\cal B}''=[{\cal
B}',{\cal B}']={\cal B}'$. These properties of ${\cal
B}=u_{\star}(n)$ and ${\cal B}'=NCsu(n)$ are analogous to those
valid for the commutative gauge algebras $u(n)$ and $su(n)$.

{\it Note:} The sub-algebra $u_{\star}(1)$ of elements (\ref{7}) with
$\epsilon(x)=\epsilon_0(x) {\bf 1}_n$ extends to another ideal
$u^n_{\star}(1)$ by adding the $[u_{\star}(1),u_{\star}(n)]$
functions, i. e. $u^n_{\star}(1)$ is formed by the elements
(\ref{7}) with $\epsilon_0(x)\in{\cal A}_{\star}$ and
$\epsilon_A(x)\in{\cal A}^{\star}_1$. Similarly, ${\bf 1}^n_{\star}\equiv 
NCsu_n(1)=[u^n_{\star}(1), u^n_{\star}(1)]$ is
an ideal in $u_{\star}(n)$.

With any ideal in $u_{\star}(n)$, we can associate the
corresponding {\it factor-algebra}: 
\be\label{8}
 u_n(1)\ := u_{\star}(n)/NCsu(n)\ ,  su(n)\ :=\ u_{\star}(n)/u^n_{\star}(1)\ . 
\ee
The ideal $u_n(1)$ is formed by
equivalency classes: $\epsilon(x)\sim \epsilon'(x)$ if
$\epsilon(x)-\epsilon'(x)\in NCsu(n)$. However, any element of
$u_{\star}(n)$ can be uniquely written in the form $\epsilon
(x)=\epsilon^0(x){\bf 1}_n +\delta^1(x)$ with $\theta$-independent
$\epsilon^0(x)\in{\cal A}$ (here ${\cal A}\equiv{\cal
A}_{\star}\setminus{\cal A}^{\star}_1$ is the factor-algebra
isomorphic to the commutative algebra of functions)
and $\delta^1(x)\in NCsu(n)$. Thus, the elements of $u_n(1)$ are
uniquely determined by the $\theta$-independent $\epsilon^0(x)$,
which themselves form, as conjugacy classes, the local Lie algebra
isomorphic to the usual commutative $u(1)$-gauge algebra.
Analogous identifications are valid for other cases in
(\ref{8}) too.

The local gauge groups $U_{\star}(n)$, $NCSU(n)$, $U^n_{\star}(1)$
and $NCSU_n(1)$ are defined by taking the star-exponent of the
corresponding ideal:
\be
U_{\star}(n)=\{\exp [i\epsilon (x)],\, \epsilon(x)\in
u_{\star}(n)\}\ ,\ NCSU(n)= \{\exp[i\epsilon (x)],\, \epsilon
(x)\in NCsu(n)\}\ , 
\ee
\be\label{9} 
U^n_{\star}(1)=\{\exp [i\epsilon (x)],\, \epsilon(x)\in u^n_{\star}(1)\}\ 
,\ 
NCSU_n(1)= \{\exp [i\epsilon (x)],\,
\epsilon (x)\in{\bf 1}^n_{\star}]\}\ . 
\ee 
The corresponding factor-groups
\be\label{10} 
U_n(1)\ =\ U_{\star}(n)/NCSU(n)\ ,  SU(n)\ =\
U_{\star}(n)/U^n_{\star}(1)\,
\ee 
are all isomorphic to the usual
gauge groups as it is indicated by the notation. Thus, we can
write: 
\be\label{11} 
U_{\star}(n)\ =\ U_n(1)\, {\bf \star}\  NCSU(n)\ =\
SU(n)\, {\bf \star}\  U^n_{\star}(1)\ .
\ee 
The meaning of the first
equality is the following: any element $U(x)\in U_{\star}(n)$ can
be written (non-uniquely) as the product $U(x)=U'(x)\star V(x)$,
where $U'(x)$ is some representant of a given conjugacy class and
some element $V(x)$ from $NCSU(n)$. Alternatively, we can consider
$U_{\star}(n)$ as a $NCSU(n)$-principal bundle over the set $U(1)$
of conjugacy classes: $NCSU(n)\to U_{\star}(n)\to U_n(1)$. More explicitly
any element of $U_*(n)$ can be uniquely written as 
\ba\label{Uxtheta}
U(x,\theta) &=&   
{\rm e}_\star^{i\epsilon_0(x){\bf 1}_n +i\epsilon_1(x,\theta){\bf
1}_n+i\epsilon_a(x,\theta)T^a} \cr
&=& {\rm e}^{i\epsilon_0(x){\bf
1}_n}\, \star\, {\rm e}_\star^{i{\tilde\epsilon}^1(x,\theta){\bf 1}_n
+i{\tilde\epsilon}_a(x,\theta)T^a}\ .
\ea 
Here ${\rm e}^{i\epsilon(x)}$ denotes the usual
exponent, whereas ${\rm e}_\star^{i\epsilon(x,\theta)}$ is the 
$\star$-exponent. 
The second factor
on R.H.S. of the second line of (\ref{Uxtheta}) is an element of the invariant 
subgroup $NCSU(n)$ being the
$\star$-exponent of elements in $NCsu(n)$. It follows that the elements of the 
factor group
$U_\star(n)/SU_\star(n)$ are uniquely specified by the first factor on
R.H.S. The product of two elements (\ref{Uxtheta}) is 
\ba\label{Uoneproduct}
\left(
{\rm e}^{i\alpha_0(x) {\bf 1}_n}\ \star {\rm e}_\star^{i{\tilde\alpha}^1_0
(x,\theta){\bf 1}_n+i{\tilde\alpha}_a(x,\theta)T^a}\right) 
&\star& 
\left({\rm e}^{i\beta_0(x){\bf 1}_n}\ \star 
{\rm e}_\star^{i{\tilde\beta}^1_0(x,\theta)
{\bf 1}_n +i{\tilde\alpha}_a(x,\theta)T^a}\right)
\cr &=&
{\rm e}^{i(\alpha_0(x)+\beta_0(x)){\bf 1}_n}\,
\star\ {\rm e}\star^{i{\tilde\gamma}^1_0(x,\theta){\bf 1}_n 
+i{\tilde\gamma}^0_c(x,\theta)T^c}\ ,
\ea
where ${\tilde\gamma}^1_0(x,\theta)$ and ${\tilde\gamma}_c(x,\theta)$
depend on all other functions appearing the L.H.S.,  
${\tilde\alpha}_A(x,\theta)$, ${\tilde\beta}_A(x,\theta)$ as well as
$\alpha_0(x)$ and $\beta_0(x)$. However, the $U(1)$ factors specifying the
elements of $U_\star (n)/SU_\star (n)$ on L.H.S. only depend on
$\alpha_0(x)$ and $\beta_0(x)$ and the $U_\star (n)/SU_\star (n)$ element 
on R.H.S. is determined by ${\rm e}^{i(\alpha_0(x)+\beta_0(x)){\bf 1}_n}$. 
We see that the factor group is {\it isomorphic} to the usual commutative local
gauge group: $U_\star(n)/NCSU(n)=U_n(1)$. We stress that in our NCSM 
construction we have only used the $U_n(1)$ and $NCSU(n)$ sub-groups.

\subsection*{Realization of the $U(1)$ gauge symmetry}
The formulas (\ref{Uxtheta}) and (\ref{Uoneproduct}) 
induce the {\it one-dimensional} representation $\pi$
of the $U_\star(n)$ group: 
\be 
\pi(U(x,\theta))\ =\ \pi(
{\rm e}_\star^{i\alpha_0(x){\bf 1}_n +i\alpha_1(x,\theta){\bf
1}_n+i\alpha_a(x,\theta)T^a})\ = \ \alpha_0(x)\ ,
\ee 
possessing the property
\be  
\pi( {\rm e}_\star^{i\alpha_0(x){\bf 1}_n +\dots}\, \star\, 
{\rm e}_\star^{i\beta_0(x){\bf 1}_n
+\dots}) = \alpha_0(x)\ +\ \beta_0(x)\ .
\ee 
This representation is realized on the gauge potentials 
\[ 
A(x,\theta)\equiv A_A(x,\theta)T^A = A_0(x){\bf 1}_n
+ A_1(x,\theta){\bf 1}_n + iA^0_a(x,\theta)T^a 
\] 
which under $U_\star(n)$ transforms in the usual way:
\be\label{gaugetransxtheta}
A(x,\theta)\ \to\ 
U(x,\theta)\star A(x,\theta)\star U^{-1}(x,\theta)\ +\ 
U(x,\theta)\star dU^{-1}(x,\theta)\ .
\ee
It can be seen that   
under (\ref{gaugetransxtheta}) the $\theta$-independent part of the gauge field 
$A_0(x)$ transforms as a usual $U(1)$ gauge field: 
\be 
A_0(x)\ \to\ A_0(x)\ +\ d\alpha_0(x)\ .
\ee

Then we can require that $\theta$-independent 
complex scalar Higgsac field $\Phi(x)$ 
under (\ref{gaugetransxtheta}) to transforms as 
\be 
\Phi\ \to\ e^{iq\alpha_o(x)}\, \Phi(x)\ ,\ q\ -\ {\rm constant}\ .
\ee  
We stress
that there are no $\star$-products on R.H.S.! An autonomous $U_n(1)$ gauge 
subsystem
can be desribed by the Higgsac action 
\be\label{Higgsacaction} 
S[A_0,\Phi]\ =\ \int dx\,
[(D(A_0)\Phi (x))^\dagger (D(A_0)\Phi(x))-V(\Phi^\dagger\Phi)]\ ,
\ee 
where $V(.,.)$ is a convienent Higgs potential and $D(A_0)=d+iqA_0(x)$ is the
$\theta$-independent $U_n(1)$ part of corresponding covariant derivative
affiliated to the (full) gauge potential $A(x,\theta)$.

Based on the the above discussions, the following notes are in order

{\it 1)} As desired, the Higgsac action $S[A_0,\Phi]$ is
$U_\star(n)$-invariant. However, the Higgsac field may interact with other   
matter fields {\it only} indirectly, via the $\theta$-independent part
$A_0(x)$ of the corresponding gauge field. 

{\it 2)} The charge $q$ of the Higgsac  field $\Phi(x)$ is
unspecified. Moreover, the Higgsac field may interact with more
($\theta$-independent parts of) gauge fields with unspecified charges.

{\it  3)} The charges are determined by constant gauge transformation
which are a part of the $\theta$-independent factor of the gauge symmetry. The
$\theta$-dependent $U_n(1)$ fields $A^1_\theta(x,\theta)$ do not feel {\it any}
$U(1)$ charge.

{\it 4)} The construction described above can be repeated for
any noncommutative associative algebra of functions possessing a
suitable filtration.

\section{Symmetry reduction}
\setcounter{equation}{0} 

Let us consider a system with
noncommutative gauge symmetry given as the direct product of two
gauge groups: 
\be\label{12} 
U_{\star}(n)\times U_{\star}(m)\ =\
(U_n(1)\times U_m(1))\, {\bf \star}\ (NCSU(n)\times NCSU(m))\ ,
\ee where the
subscripts of the $U(1)$ factors indicate to which
$U_{\star}(\cdot)$ gauge group they belong.

The symmetry reduction consists of replacing the two independent
$U(1)$ factors  
\ba\label{13} 
U_n(1) &=& \{\exp [{i\over 2}\epsilon^0_n(x){\bf
1}_n]\}\ ,\ \epsilon^0_n(x)\in{\cal A}\ , \cr
U_m(1)&=& \{\exp [{i\over 2}\epsilon^0_m(x){\bf 1}_m]\}\ ,\
\epsilon^0_m(x)\in{\cal A}\ , 
\ea 
by a ``diagonal'' one specified by
putting $\epsilon^0_n(x)={1\over n}\epsilon^0(x)$ and
$\epsilon^0_m(x)={1\over m}\epsilon^0(x)$ with
$\epsilon^0(x)\in{\cal A}$: 
\ba\label{14} 
(U_n(1)\times U_m(1))_d
&\equiv &\ \{\exp [{i\over 2}({1\over n}{\bf 1}_n \oplus {1\over
m}{\bf 1}_m)\epsilon^0(x)]\} \cr
 &=& \{\exp[{i\over 2n}\epsilon^0(x){\bf 1}_n]\} \times \{\exp{i\over
2m}\epsilon^0(x){\bf 1}_m]\}\ ,
\ea
 where $\epsilon^0(x)\in{\cal
A}$ and the symbol $\oplus$ denotes the direct sum. In $U_n(1)$
we can introduce the determinat by: ${\rm det}(\exp [{i\over
2}\epsilon^0_n(x){\bf 1}_n])=\exp [{i\over 2}n\epsilon^0_n(x)]$.
The ${1\over n}$-factors guarantee that ${\rm det}(\exp [{i\over
2n}\epsilon^0(x){\bf 1}_n])={\rm det}(\exp [{i\over
2m}\epsilon^0(x){\bf 1}_m])=\exp [{i\over 2}\epsilon^0 (x)]$ is a
representation of $(U_n(1)\times U_m(1))_d$. After the symmetry
reduction we are left with the gauge group
\ba\label{15} 
(U_{\star}(n)\times U_{\star}(m))_d\ &\equiv & (U_n(1)\times U_m(1))_d\, 
{\bf \star}\  (NCSU(n)\times NCSU(m)) \cr 
&= & (U_{\star}(n)\times
U_{\star}(m))/({\rm det}({U_n}(1))_d={\rm det}(U_m(1))_d .
\ea 
In other
words, the gauge groups $NCSU(n)$ and $NCSU(m)$ are not
supplemented by two independent factors $U_n(1)$ and $U_m(1)$ but
only by one diagonal factor $(U_n(1)\times U_m(1))_d$ containing
strictly related factors $(U_n(1))_d$ and $(U_m(1))_d$ with equal
determinants.

In the language of gauge fields it means the following:
Originally, we have two gauge fields $A^n_\mu (x)$ and $A^m_\mu
(x)$  sharing the gauge transformations of
$U_n(1)=U_{\star}(n)/NCSU(n)$ and $U_m(1)=U_{\star}(m)/NCSU(m)$:
they can be identified with the $\theta$-independent parts of
${\bf 1}_n$ and ${\bf 1}_m$ components of $U_{\star}(n)$ and
$U_{\star}(m)$ gauge fields, respectively. Under the $U_n(1)$ and
$U_m(1)$ gauge transformations they transform as 
\ba\label{16} 
A^n_\mu (x)
&\to & A^n_\mu (x)+{i\over 2}g^{-1}_n\partial_\mu
\epsilon^0_n(x){\bf 1}_n\  ,\ \epsilon^0_n(x)\in{\cal A}\ , \cr
A^m_\mu (x) &\to & A^m_\mu (x)+{i\over 2}g^{-1}_m
\partial_\mu\epsilon^0_m(x){\bf 1}_m\ ,\ \epsilon^0_m(x)\in{\cal A}\ .
\ea
After the symmetry reduction  (\ref{13})-(\ref{14}) there is one
$\theta$-independent gauge field $A^d_\mu(x)$ sharing the
$(U_n(1)\times U_m(1))_d$ gauge symmetry: 
\be\label{17}
A^d_\mu(x)\ \equiv\ (A^n_\mu(x)\oplus A^m_\mu(x))_d\ =\ ({g\over
ng_n}{\bf 1}_n\oplus {g\over mg_m}{\bf 1}_m)A_\mu (x)\ .
\ee 
The $\theta$-independent field $A_\mu (x)$ transforms under
$(U_n(1)\times U_m(1))_d$ gauge transformations generated by
$\epsilon^0 (x)$ as follows: 
\be\label{18} 
A_\mu (x)\ \to\ A_\mu
(x)+{i\over 2}g^{-1}\partial_\mu\epsilon^0 (x)\ .
\ee 
It is important that 
(\ref{16}) coincides on $(U_n(1)\times U_m(1))_d$
transformations with (\ref{17})-(\ref{18}). The constant $g$ is
specified by the equation: 
\be\label{19} 
{1\over g^2}\ =\ {1\over n^2g^2_n}\ + {1\over m^2g^2_m} 
\ee 
(this guarantees the proper normalization of the $A_\mu$-field term in 
Lagrangian).

\section{Symmetry reduction, fermionic part}
\setcounter{equation}{0} 
In NCSM the symmetry reduction is
mediated by a $\theta$-independent Higgsac field $\Phi (x)\in{\cal
A}$ possessing $(U_n(1)\times U_m(1))$ transformations: 
$$ 
\Phi(x)\ \to\ u(x)\Phi (x)v^{-1}(x)\ =\ \Phi (x)\ , 
$$ 
$$ 
u(x)={\rm det}\exp [{i\over 2}\epsilon^0_n(x){\bf 1}_n]\ ,\ v(x)={\rm
det}\exp [{i\over 2}\epsilon^0_m(x){\bf 1}_m]\ . 
$$ 
Note that there is no $\star$-product invovled in the gauge transformation of 
the Higgsac field.
However, for $(U_n(1)\times U_m(1))_d$ transformations: 
$\Phi (x)\to\Phi (x)$
(since $\epsilon^0_n(x)={1\over n}\epsilon^0(x)$, $\epsilon^0_m(x)
={1\over m}\epsilon^0(x)$, and both phase factors cancel). Thus,
the Higssac field is {\it neutral} with respect to the residual
gauge field $A_\mu (x)$. This is consistent with the observation
that the $\Phi (x)$ field covariant derivative ($\partial_\mu
+ig_n{\rm det}A^n_\mu (x)-ig_m{\rm det}A^m_\mu (x)$) does not
transform under $(U_n(1)\times U_m(1))_d$ gauge transformations.

Let us now consider the matter fields $\Psi_u(x)$ and $\Psi_d(x)$
transforming under $U_{\star}(n)\times U_{\star}(m)$ gauge
transformation as follows: 
\be\label{20} 
\Psi_u(x)\ \to\ U(x)\Psi_u(x)V^{-1}(x)\ ,\ 
\Psi_d(x)\ \to\ \Psi_d(x)V^{-1}(x)\ ,
\ee 
with $U(x)\in U_{\star}(n)$ and $V(x)\in U_{\star}(m)$. Their
$NCSU(n)\times NCSU(m)$ orbits are: 
$$ 
\{U(x)\Psi_u(x)V^{-1}(x),\
U(x)\in NCSU(n),\, V(x)\in NCSU(m)\}\ ,
$$ 
\be\label{21} \{
\Psi_d(x)V^{-1},\ V(x)\in NCSU(m)\}\ , 
\ee 
This means that if
$\psi_u(x)$ and $\psi_d(x)$ are representatives of the classes in
question, then 
\be\label{22} 
\Psi_u(x)\ =\ U(x)\psi_u(x)V^{-1}(x)\ ,\ 
\Psi_d(x)\ =\ \psi_d(x){V}^{-1}(x)\ 
,\ee 
with $U(x)\in NCSU(n)$ and $V(x)\in NCSU(m)$. The fields $\psi_u(x)$ and
$\psi_d(x)$ transform under the $U_n(1)\times U_m(1)$ gauge
transformation (\ref{16}) as follows: 
\ba\label{23} 
\psi_u(x)\ &\to\ & \exp
[{i\over 2}\epsilon^0_n(x){\bf 1}_n]\psi_u(x)\exp [-{i\over
2}\epsilon^0_m(x){\bf 1}_m]\ ,\cr
\psi_d(x)\ &\to\ &
\psi_d(x)\exp [-{i\over 2}\epsilon^0_m(x){\bf 1}_m]\ . 
\ea
Restricting (\ref{23}) to $(U_n(1)\times U_m(1))_d$
transformations by putting $\epsilon^0_n(x)={1\over
n}\epsilon^0(x)$ and $\epsilon^0_m(x)={1\over m}\epsilon^0(x)$ we
obtain that the orbits transform as: 
\ba\label{24} 
\psi_u(x)\ &\to\ & \exp
[{i\over 2n}\epsilon^0(x){\bf 1}_n]\, \psi_u(x)\, \exp [-{i\over
2m}\epsilon^0(x){\bf 1}_m]\ ,\cr
\psi_d(x)\ & \to\ &
\psi_d(x)\, \exp [-{i\over 2m}\epsilon^0(x){\bf 1}_m] \ .
\ea
Comparing this with (\ref{18}) we see that they possess fractional
$A_\mu$-field charges: 
\be\label{25} 
q_u\ =\ {g\over n}-{g\over
m}\ ,\ q_d\ =\ -{g\over m}\ .
\ee

This is the solution of the {\it fractional charge} mystery in NC
QFT: they appear as charges of the $\theta$-independent residual
gauge field $A_\mu (x)$ which transforms like a commutative $U(1)$
gauge field and can interact with the fields $\psi_u(x)$ and
$\psi_d(x)$ possessing fractional charges. We can extend this as
follows. The fields $\Psi_u(x)$ and $\Psi_d(x)$ themselves, and
not only $\psi_u(x)$ and $\psi_d(x)$, possess the fractional
charges $q_u$ and $q_d$ given above. This is reasonable, since
charges are determined by global transformations with
$\epsilon^0(x) =const$, and eqs. (\ref{24}) and (\ref{25}) with
constant $\epsilon^0(x)$ are valid directly for the matter fields
$\Psi_u(x)$ and $\Psi_d(x)$. The exact values of charges given in
(\ref{22}) can be read directly from the the field transformation
law (\ref{22}). This simple rule is valid for any field.

{\it Note}: These are exactly the formulas discussed in the
sections 3 and 4. For example, the formula for the charge of
$\Psi_u(x)$ given there ($q_u={1\over 2}(g_n\cos
\delta_{nm}-g_m\sin\delta_{nm})$, $\tan\delta_{nm}=ng_n/mg_m$) is
identical to (\ref{25}). However, the motivation presented here is
``kinematical'', being based only on symmetry considerations, the
symmetry reducing part of the Lagrangian is not specified. We see
that the symmetry reduction is related only to the
$\theta$-independent parts of fields sharing the corresponding
commutative $U_n(1)\times U_m(1)$ factor-group symmetries.

\end{document}